\documentclass[]{IEEEtran}

\usepackage{mathtools, amssymb, amsthm}
    \def\le{\leqslant}
    \def\ge{\geqslant}
    \def\EE{\mathbb E}
    \def\BB{\mathcal B}
    \def\II{\mathcal I}
    \def\BDC{\operatorname{BDC}}
    \def\BSC{\operatorname{BSC}}
    \def\QSC{\operatorname{QSC}}
    \newtheorem{theorem}{Theorem}
    \newtheorem{problem}[theorem]{Problem}

\usepackage{tikz, booktabs}
    \tikzset{every picture/.style={line cap=round, line join=round}}

\usepackage[colorlinks=true, allcolors=blue!50!purple]{hyperref}

\begin{document}

\title{Block Length Gain for Nanopore Channels}
\author{Yu-Ting Lin, Hsin-Po Wang, Venkatesan Guruswami}

\maketitle

\begin{abstract}
    DNA is an attractive candidate for data storage.
    Its millennial durability and nanometer scale
    offer exceptional data density and longevity.
    Its relevance to medical applications also drives
    advances in DNA-related biotechnology.

    To protect our data against errors, a straightforward approach
    uses one error-correcting code per DNA strand,
    with a Reed--Solomon code protecting the collection of strands.
    A downside is that current technology can only
    synthesize strands 200--300 nucleotides long.
    At this block length, the inner code rate suffers a significant
    finite-length penalty, making its effective capacity hard to characterize.

    Last year, we proposed \emph{Geno-Weaving}
    in a JSAIT publication.
    The idea is to protect the same position
    across multiple strands using one code;
    this provably achieves capacity against substitution errors.
    In this paper, we extend the idea to combat deletion errors
    and show two more advantages of Geno-Weaving:
    (1) Because the number of strands is 3--4 orders of magnitude larger
        than the strand length, the finite-length penalty vanishes.
    (2) At realistic deletion rates $0.1\%$--$10\%$,
        Geno-Weaving designed for BSCs works well empirically,
        bypassing the need to tailor the design for deletion channels.
\end{abstract}

\begin{figure*}
    \tikzset{
        base/.style={
            transform shape, above right,
            circle, draw, fill, inner sep=1pt,
        }
    }
    \begin{tikzpicture} [baseline=0]
        \foreach \x in {1, 2, 3} {
            \tikzset{shift = {(\x, 0)}}
            \draw [fill=.!10] (0, 0) rectangle (0.1, 0.5);
            \draw
                (-0.1, 0.4) -- (-0.1, 0.6) -- (0.2, 0.6) -- (0.2, 0.4)
                (0.05, 0.6) node [above] {site\x}
            ;
        }
        \draw (0.9, 0) -- node [below] {\strut sites with caps} (3.5, 0);
    \end{tikzpicture}
    \hfill
    \begin{tikzpicture} [baseline=0]
        \foreach \x in {1, 2, 3} {
            \tikzset{shift = {(\x, 0)}}
            \draw [fill=.!10] (0, 0) rectangle (0.1, 0.5);
            \ifnum \x = 2
                \draw [dash pattern=on0.1pt off0.8pt]
                    {[rotate=20] (-0.1, 0.4) -- (-0.1, 0.6)}
                    {[rotate=10, yshift=2] (-0.1, 0.6) -- (0.2, 0.6)}
                    {[rotate=-5] (0.2, 0.6) -- (0.2, 0.4)}
                ;
            \else
                \draw (-0.1, 0.4) -- (-0.1, 0.6) -- (0.2, 0.6) -- (0.2, 0.4);
            \fi
        }
        \draw [red, line width=1pt] (2.1, 0.5) circle (1pt) -- +(0.3, 3);
        \draw
            (0.9, 2.5) -- node [below] {mask}
            (2.2, 2.5) (2.4, 2.5) -- (3.5, 2.5)
        ;
        \foreach \x in {1, 1.5, 2.5, 3} {
            \draw [red, line width=1pt]
                (\x+.3, 2.5) circle (1pt) -- +(0.1, 1) ;
        }
        \draw (0.9, 0) -- node [below] {\strut light breaks caps} (3.5, 0);
    \end{tikzpicture}
    \hfill
    \begin{tikzpicture} [baseline=0]
        \foreach \x in {1, 2, 3} {
            \tikzset{shift = {(\x, 0)}}
            \draw [fill=.!10] (0, 0) rectangle (0.1, 0.5);
            \ifnum \x = 2 \else
                \draw (-0.1, 0.4) -- (-0.1, 0.6) -- (0.2, 0.6) -- (0.2, 0.4);
            \fi
        }
        \foreach \x in {1, 2, 3} {
            \foreach \y in {1, 2, 3, 4}{
                \tikzset{
                    overlay,
                    shift = {({\x+rand*0.1 + 0.2*(-1)^\y}, \y*0.7+rand*0.1)},
                    rotate around={rand*180:(0.2,0.3)}
                }
                \draw [blue, fill=.!10]
                    (0, 0.5) rectangle (0.1, 0) node [base] {A};
                \draw [blue]
                    (-0.1, 0.4) -- (-0.1, 0.6) -- (0.2, 0.6) -- (0.2, 0.4);
            }
        }
        \draw (0.9, 0) -- node [below] {\strut inject letter A} (3.5, 0);
    \end{tikzpicture}
    \hfill
    \begin{tikzpicture} [baseline=0]
        \foreach \x in {1, 2, 3} {
            \tikzset{shift = {(\x, 0)}}
            \draw [fill=.!10] (0, 0) rectangle (0.1, 0.5);
            \ifnum \x = 2 \else
                \draw
                    (-0.1, 0.4) -- (-0.1, 0.6) -- (0.2, 0.6) -- (0.2, 0.4);
            \fi
        }
        \foreach \x in {1, 2, 3} {
            \foreach \y in {3, 4}{
                \tikzset{
                    overlay,
                    shift = {({\x+rand*0.1 + 0.2*(-1)^\y}, \y*0.7+rand*0.1)},
                    rotate around={rand*180:(0.2,0.3)}
                }
                \draw [blue, fill=.!10]
                    (0, 0.5) rectangle (0.1, 0) node [base] {A};
                \draw [blue]
                    (-0.1, 0.4) -- (-0.1, 0.6) -- (0.2, 0.6) -- (0.2, 0.4);
            }
        }
        \draw [shift = {(2, 0.5)}, blue, fill=.!10]
            (0, 0.5) rectangle (0.1, 0) node [base] {A};
        \draw [shift = {(2, 0.5)}, blue]
            (-0.1, 0.4) -- (-0.1, 0.6) -- (0.2, 0.6) -- (0.2, 0.4);
        \draw (0.9, 0) -- node [below] {\strut A attaches to it} (3.5, 0);
    \end{tikzpicture}
    \hfill
    \begin{tikzpicture} [baseline=0]
        \foreach \x in {1, 2, 3} {
            \tikzset{shift = {(\x, 0)}}
            \draw [fill=.!10] (0, 0) rectangle (0.1, 0.5);
            \ifnum \x = 1
                \draw
                    (-0.1, 0.4) -- (-0.1, 0.6) -- (0.2, 0.6) -- (0.2, 0.4);
            \fi
        }
        \draw [red, line width=1pt] (2.1, 1) circle (1pt) -- +(0.25, 2.5);
        \draw [red, line width=1pt] (3.1, 0.5) circle (1pt) -- +(0.3, 3);
        \draw
            (0.9, 2.5) -- node [below] {mask}
            (2.15, 2.5) (2.35, 2.5) -- (3.2, 2.5) (3.4, 2.5) -- (3.5, 2.5)
        ;
        \foreach \x in {1, 1.5, 2.5} {
            \draw [red, line width=1pt]
                (\x+.3, 2.5) circle (1pt) -- +(0.1, 1);
        }
        \draw [blue, fill=.!10, shift = {(2, 0.5)}]
            (0, 0.5) rectangle (0.1, 0) node [base] {A};
        \draw [dash pattern=on0.1pt off0.8pt, shift = {(2, 0.5)}, blue]
            {[rotate=20] (-0.1, 0.4) -- (-0.1, 0.6)}
            {[rotate=10, yshift=2] (-0.1, 0.6) -- (0.2, 0.6)}
            {[rotate=-5] (0.2, 0.6) -- (0.2, 0.4)}
        ;
        \draw [dash pattern=on0.1pt off0.8pt, shift = {(3, 0)}]
            {[rotate=20] (-0.1, 0.4) -- (-0.1, 0.6)}
            {[rotate=10, yshift=2] (-0.1, 0.6) -- (0.2, 0.6)}
            {[rotate=-5] (0.2, 0.6) -- (0.2, 0.4)}
        ;
        \draw (0.9, 0) -- node [below] {\strut break more caps} (3.5, 0);
    \end{tikzpicture}
    \hfill
    \begin{tikzpicture} [baseline=0]
        \foreach \x in {1, 2, 3} {
            \tikzset{shift = {(\x, 0)}}
            \draw (0, 0) rectangle (0.1, 0.5);
            \ifnum \x = 1
                \draw (-0.1, 0.4) -- (-0.1, 0.6) -- (0.2, 0.6) -- (0.2, 0.4);
            \fi
        }
        \foreach \x in {1, 2, 3} {
            \foreach \y in {3, 4}{
                \tikzset{
                    overlay,
                    shift = {({\x+rand*0.1 + 0.2*(-1)^\y}, \y*0.7+rand*0.1)},
                    rotate around={rand*180:(0.2,0.3)}
                }
                \draw [teal, fill=.!10]
                    (0, 0.5) rectangle (0.1, 0) node [base] {C} (0.5, 0.4);
                \draw [teal]
                    (-0.1, 0.4) -- (-0.1, 0.6) -- (0.2, 0.6) -- (0.2, 0.4);
            }
        }
        \draw [blue, fill=.!10, shift = {(2, 0.5)}]
            (0, 0.5) rectangle (0.1, 0) node [base] {A};
        \draw [teal, fill=.!10, shift = {(2, 1)}]
            (0, 0.5) rectangle (0.1, 0) node [base] {C};
        \draw [teal, fill=.!10, shift = {(3, 0.5)}]
            (0, 0.5) rectangle (0.1, 0) node [base] {C};
        \draw [teal, shift = {(2, 1)}]
            (-0.1, 0.4) -- (-0.1, 0.6) -- (0.2, 0.6) -- (0.2, 0.4);
        \draw [teal, shift = {(3, 0.5)}]
            (-0.1, 0.4) -- (-0.1, 0.6) -- (0.2, 0.6) -- (0.2, 0.4);
        \draw (0.9, 0) -- node [below] {\strut this time inject C} (3.5, 0);
    \end{tikzpicture}
    \caption{
        The photolithographic approach to synthesize arbitrary DNA.
        From left to right:
        Prepare multiple \emph{sites} on a base plate,
        each protected by a cap.
        Selectively break the caps for the sites we want to extend;
        this is done by lighting, heating, or charging.
        Inject the next letter we want to extend the sites by
        (A in this case).
        The A's naturally attach to sites without a cap.
        Afterwards, break other caps and inject other letters
        to extend the sites further.
    }                                                         \label{fig:syn}
\end{figure*}
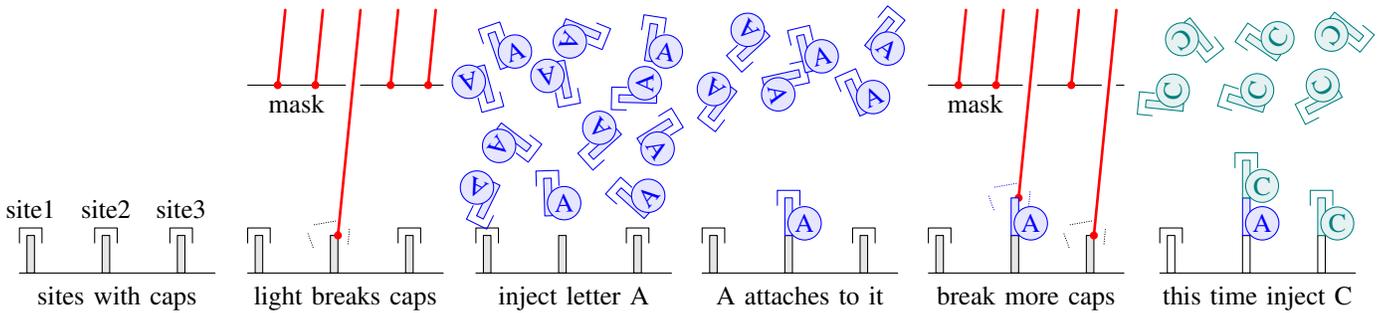

\section{Introduction}

    DNA is a tiny molecule that evolution selected
    to carry genetic information.
    At its scale, it is hard to imagine a competitor
    in terms of stability and reliability:
    each of A, T, C, and G is about 0.34nm long along the backbone
    while modern CPUs\footnote{SSDs use even larger gates.} use 2nm gates.
    Meanwhile, ancient DNA
    can be recovered from polar zones after two million years \cite{KWD22},
    unlike modern storage media that last for decades\footnote{
    They are \emph{designed} to last for decades.
    They may last longer, but it is hard to tell for technologies
    less than a century old.} \cite{Bye03}.
    These properties make DNA an interesting data storage medium.

    In the last two decades, the bio industry has pushed toward
    better DNA sequencing \cite{YHZ22} and synthesis \cite{HVS23},
    the former for estimating the probability of catching cancers
    and the latter for customizing genes and proteins.
    Investment has driven down the cost of reading and writing DNA
    by orders of magnitude.
    Though the future is hard to predict, it is reasonable
    to expect that synthesis and sequencing will continue to improve.
    Practical DNA data storage may soon reach the market,
    starting with niche applications such as watermarking
    and store-now/decrypt-later.
    
    DNA's unique properties also pose
    interesting challenges for coding theorists.
    One that is brought up fairly often is the nanopore channel model.
    The \emph{nanopore sequencer} \cite{YHZ22} is a hand-held, low-cost,
    real-time DNA sequencing device;
    it reads DNA floating in an electrolyte solution
    by driving it through a nanoscale pore and measuring current.
    Now,
    \begin{itemize}
        \item Because strands float freely in the pool,
            they will be read out of order.
        \item Because strands are amplified,
            each one will be sampled a random number of times.
        \item Electrical signals must be converted to ACGT strings,
            introducing substitution, deletion, repetition, and insertion
            errors.
    \end{itemize}
    Note that each bullet on its own is a recurrent challenge
    that has been studied before for other reasons.
    For instance, internet packets are also received out of order
    because they travel different routes.
    But because the Internet backbone has error correction built in,
    missing packets (i.e., erasures) constitute the majority of errors.
    In contrast, the nanopore channel experiences a multitude of errors,
    including generic substitution errors and synchronization errors
    (deletions, insertions, and repetition).
    The latter are notoriously difficult to handle.

    According to the survey \cite{SKS24},
    coding theorists have worked tirelessly
    to mitigate deletions and insertions in the past decade.
    To begin, the classical result of Levenshtein \cite{Lev66} states that
    recovering from $d$ deletions and $i$ insertions
    can be reduced to recovering from $d + i$ deletions.
    This makes deletion channel the primary target in this community.
    Next, we have seen code constructions that targeted
    what we called \emph{combinatorial deletion channels},
    i.e., the channels that delete a fixed number of symbols per block.
    Finally, more works started targeting the capacity of
    \emph{probabilistic deletion channels},
    i.e., each symbol is deleted independently with a fixed probability.
    upper bounds, lower bounds, and asymptotic behaviors are given,
    as well as code constructions that approach the capacity.

    The development of combinatorial and probabilistic deletion codes
    can be applied to DNA coding
    via the following concatenation-based framework:
    Protect the data by applying one deletion code to each DNA strand.
    Given that strands might be missing or decoded incorrectly,
    apply an outer code, such as a Reed--Solomon code \cite{OAC18},
    that treats each strand as a symbol over a (very) large alphabet.
    This framework has several limitations.
    We discuss three below.

    \textbf{(A) Block length penalty.}
    Even if a code design can approach the capacity of
    a deletion channel (probabilistic or combinatorial),
    it usually requires a long block length to mitigate some overhead.
    Current DNA synthesis technology
    limits the strand length to a few hundreds of nucleotides,
    beyond which the error rate increases significantly.
    At such short lengths, the finite-length penalty is
    \emph{provably} significant.

    \textbf{(B) Fast-fading nature.}
    The fact that nanopores sample DNA strands
    implies that some strands might be read multiple times,
    while some strands might be missing altogether,
    and there is no way to predict\footnote{
    Note that this issue has very little to do with poor channel estimation:
    the number of reads per strand is not a fixed number
    that we need to estimate accurately before encoding,
    it is a random number that is drawn independently for each strand
    after the encoding is done.} the multiplicity at the encoder side.
    Therefore, a perfect code withstanding $2$ deletions
    becomes helpless if the strand is read only once with $3$ deletions,
    but wasteful if the strand is read $5$ times
    and some simple majority voting corrects all $3$ deletions.

    \textbf{(C) Goal misalignment.}
    The medical origin of DNA sequencing technology bears a hidden
    assumption that all symbols should be recognized correctly at all cost,
    as one wrong base in a gene can totally change a protein.
    This, however, aligns poorly with coding theory's theme
    that a small fraction of errors can be tolerated because
    we get to choose, and will add redundancy into, the data.
    Consequently, it is reasonable to study channels noisier than what
    current technology offers because we might want to use a cheaper device
    in a less controlled environment.
    
    Geno-Weaving \cite{WaG25} was developed to address the limitations above.
    It tries to bypass (A)'s finite-length penalty
    by coding in the \emph{orthogonal} direction,
    the direction where the strand index $s \in [n]$ varies
    while the position index $p \in [\ell]$ is fixed.
    Typically, the number of strands $n$,
    is in the order of millions \cite{OAC18},
    while the strand length $\ell$ is a few hundreds.
    This solves (A) immediately.
    Coding in the orthogonal direction also helps with (B)
    because the number of reads per strand
    can be modeled\footnote{
    As an example, consider a channel that will feed the input $X$ into
    $K$ iid copies of BSC$(1\%)$ and output the channels' outputs
    as a $K$-tuple,
    where $K$ follows the Poisson distribution with intensity $1$.
    This is a discrete channel with output alphabet
    $\bigcup_{k\ge0} \{0, 1\}^k$ and we have theories (e.g., \cite{STA09})
    on how to construct good error-correcting codes over it.}
    as part of the iid channel randomness.
    Finally, since Geno-Weaving uses polar codes,
    it can handle any substitution channel, even non-symmetric ones,
    and resolves (C).

    One aspect not covered in \cite{WaG25} is insertion and deletion errors.
    Part of the reason is that the substitution model has a clearer
    capacity expression and code design is easier
    given the abundance of available tools.
    In this paper, we try to shrink the gap
    by applying Geno-Weaving to nanopore channels with deletions.
    We provide theoretical analysis and simulation results,
    both showing that Geno-Weaving outperforms the concatenation approach.
    Moreover, readers will see that
    we do not need to study deletion channels to obtain a good code.
    This is because our coding direction is orthogonal to
    the desynchronizing direction,
    once again demonstrating the strength of Geno-Weaving.

    We organize the paper as follows.
    Section~\ref{sec:bio} reviews current biotechnology
    and states the nanopore channel model used in this work.
    Section~\ref{sec:concat} overviews available code designs
    for deletion channels and discusses implied capacity on nanopore channels.
    Section~\ref{sec:weave} recaps Geno-Weaving
    and estimates its capacity on nanopore channels.
    Finally, Section~\ref{sec:sim} presents simulation results.

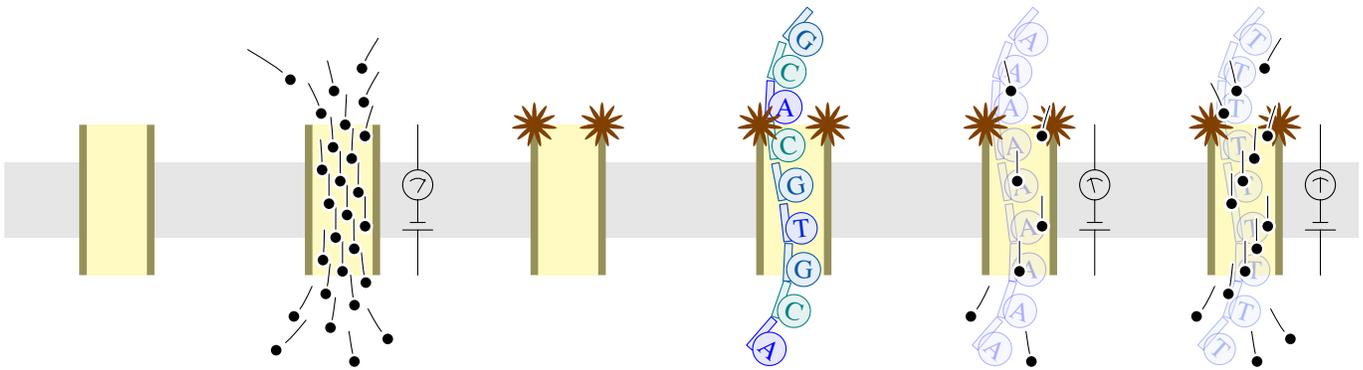
\begin{figure*}
    \centering
    \tikzset{
        A/.style={blue, fill=.!10},
        C/.style={teal, fill=.!10},
        G/.style={blue!30!teal, fill=.!10},
        T/.style={blue!70!teal, fill=.!10},
        base/.style={
            transform shape, above right,
            circle, draw, fill, inner sep=1pt,
        }
    }
    \begin{tikzpicture}
        \fill [.!10] (0, -0.5) rectangle (18, 0.5);

        \tikzset{shift = {(1.5, 0)}}
        
        \fill [yellow!30] (-0.4, 1) rectangle (0.4, -1);
        \fill [yellow!50!black]
            (-0.5, 1) rectangle (-0.4, -1) (0.4, 1) rectangle (0.5, -1);
        
        \tikzset{shift = {(3, 0)}}
        
        \fill [yellow!30] (-0.4, 1) rectangle (0.4, -1);
        \fill [yellow!50!black]
            (-0.5, 1) rectangle (-0.4, -1) (0.4, 1) rectangle (0.5, -1);
        \foreach \k in {-13, ..., 13} {
            \pgfmathsetmacro\y{\k * 0.15} 
            \pgfmathsetmacro\ya{\y + 0.2} 
            \pgfmathsetmacro\yb{\y + 0} 
            \pgfmathsetmacro\yc{\y - 0.1} 
            \pgfmathsetmacro\yd{\y - 0.2} 
            \pgfmathsetmacro\x{(mod(999 + \k*13, 21) - 10) * 0.03}
            \pgfmathsetmacro\xa{\x * (1 + (\ya)^4 * 0.2)}
            \pgfmathsetmacro\xb{\x * (1 + (\yb)^4 * 0.2)}
            \pgfmathsetmacro\xc{\x * (1 + (\yc)^4 * 0.2)}
            \pgfmathsetmacro\xd{\x * (1 + (\yd)^4 * 0.2)}
            \draw [preaction={draw=white, line width=2pt}]
                (\xa, \ya) .. controls (\xb, \yb) .. (\xc, \yc);
            \fill [preaction={draw=white, line width=2pt}]
                (\xd, \yd) circle (2pt);
        }
        \draw [shift = {(1, 0)}]
            (0, 1) -- (0, 0.4)
            (0, 0.2) circle (0.2)
            (0, 0.1) -- +(60:0.2)
            (-0.1, 0.25) to [bend left=40] (0.1, 0.25)
            (0, 0) -- (0, -0.3)
            (-0.1, -0.3) -- (0.1, -0.3)
            (-0.2, -0.4) -- (0.2, -0.4)
            (0, -0.4) -- (0, -1)
        ;
        
        \tikzset{shift = {(3, 0)}}
        
        \fill [yellow!30] (-0.4, 1) rectangle (0.4, -1);
        \fill [yellow!50!black]
            (-0.5, 1) rectangle (-0.4, -1) (0.4, 1) rectangle (0.5, -1);
        \foreach \t in {1, ..., 11} {
            \fill [shift = {(-0.45, 1)}, rotate=\t*360/11, orange!50!black]
                (0, 0.1) ellipse (0.03 and 0.2);
            \fill [shift = {(0.45, 1)}, rotate=\t*360/11, orange!50!black]
                (0, 0.1) ellipse (0.03 and 0.2);
        }

        \tikzset{shift = {(3, 0)}}
        
        \fill [yellow!30] (-0.4, 1) rectangle (0.4, -1);
        \fill [yellow!50!black]
            (-0.5, 1) rectangle (-0.4, -1) (0.4, 1) rectangle (0.5, -1);
        \foreach \A [count=\k] in {A, C, G, T, G, C, A, C, G} {
            \pgfmathsetmacro\y{\k - 5}
            \pgfmathsetmacro\X{\y^3 * 0.01 - \y * 0.1 - 0.25}
            \pgfmathsetmacro\Y{\y * 0.55 * (1 - (\y)^2 * 0.004)}
            \pgfmathsetmacro\an{-(\y)^2 * 3.2 + 10 + rand}
            \tikzset{shift = {(\X, \Y)}, rotate around= {\an : (0.1, 0.25)}}
            \draw [\A]
                (0, 0.5) rectangle (0.1, 0) node [base] {\A} (0.5, 0.4);
        }
        \foreach \t in {1, ..., 11} {
            \fill [shift = {(-0.45, 1)}, rotate=\t*360/11, orange!50!black]
                (0, 0.1) ellipse (0.03 and 0.2);
            \fill [shift = {(0.45, 1)}, rotate=\t*360/11, orange!50!black]
                (0, 0.1) ellipse (0.03 and 0.2);
        }
        
        \tikzset{shift = {(3, 0)}}
        
        \fill [yellow!30] (-0.4, 1) rectangle (0.4, -1);
        \fill [yellow!50!black]
            (-0.5, 1) rectangle (-0.4, -1) (0.4, 1) rectangle (0.5, -1);
        \foreach \A [count=\k] in {A, A, A, A, A, A, A, A, A} {
            \pgfmathsetmacro\y{\k - 5}
            \pgfmathsetmacro\X{\y^3 * 0.01 - \y * 0.1 - 0.25}
            \pgfmathsetmacro\Y{\y * 0.55 * (1 - (\y)^2 * 0.004)}
            \pgfmathsetmacro\an{-(\y)^2 * 3.2 + 10 + rand}
            \tikzset{shift = {(\X, \Y)}, rotate around= {\an : (0.1, 0.25)}}
            \draw [\A, opacity=0.3]
                (0, 0.5) rectangle (0.1, 0) node [base] {\A} (0.5, 0.4);
        }
        \foreach \t in {1, ..., 11} {
            \fill [shift = {(-0.45, 1)}, rotate=\t*360/11, orange!50!black]
                (0, 0.1) ellipse (0.03 and 0.2);
            \fill [shift = {(0.45, 1)}, rotate=\t*360/11, orange!50!black]
                (0, 0.1) ellipse (0.03 and 0.2);
        }
        \foreach \k in {-13, -9, ..., 13} {
            \pgfmathsetmacro\y{\k * 0.15} 
            \pgfmathsetmacro\ya{\y + 0.2} 
            \pgfmathsetmacro\yb{\y + 0} 
            \pgfmathsetmacro\yc{\y - 0.1} 
            \pgfmathsetmacro\yd{\y - 0.2} 
            \pgfmathsetmacro\x{(mod(999 + \k*13, 21) - 10) * 0.03}
            \pgfmathsetmacro\xa{\x * (1 + (\ya)^4 * 0.2)}
            \pgfmathsetmacro\xb{\x * (1 + (\yb)^4 * 0.2)}
            \pgfmathsetmacro\xc{\x * (1 + (\yc)^4 * 0.2)}
            \pgfmathsetmacro\xd{\x * (1 + (\yd)^4 * 0.2)}
            \draw [preaction={draw=white, line width=2pt}]
                (\xa, \ya) .. controls (\xb, \yb) .. (\xc, \yc);
            \fill [preaction={draw=white, line width=2pt}]
                (\xd, \yd) circle (2pt);
        }
        \draw [shift = {(1, 0)}]
            (0, 1) -- (0, 0.4)
            (0, 0.2) circle (0.2)
            (0, 0.1) -- +(105:0.2)
            (-0.1, 0.25) to [bend left=40] (0.1, 0.25)
            (0, 0) -- (0, -0.3)
            (-0.1, -0.3) -- (0.1, -0.3)
            (-0.2, -0.4) -- (0.2, -0.4)
            (0, -0.4) -- (0, -1)
        ;
        
        \tikzset{shift = {(3, 0)}}
        
        \fill [yellow!30] (-0.4, 1) rectangle (0.4, -1);
        \fill [yellow!50!black]
            (-0.5, 1) rectangle (-0.4, -1) (0.4, 1) rectangle (0.5, -1);
        \foreach \A [count=\k] in {T, T, T, T, T, T, T, T, T} {
            \pgfmathsetmacro\y{\k - 5}
            \pgfmathsetmacro\X{\y^3 * 0.01 - \y * 0.1 - 0.25}
            \pgfmathsetmacro\Y{\y * 0.55 * (1 - (\y)^2 * 0.004)}
            \pgfmathsetmacro\an{-(\y)^2 * 3.2 + 10 + rand}
            \tikzset{shift = {(\X, \Y)}, rotate around= {\an : (0.1, 0.25)}}
            \draw [\A, opacity=0.3]
                (0, 0.5) rectangle (0.1, 0) node [base] {\A} (0.5, 0.4);
        }
        \foreach \t in {1, ..., 11} {
            \fill [shift = {(-0.45, 1)}, rotate=\t*360/11, orange!50!black]
                (0, 0.1) ellipse (0.03 and 0.2);
            \fill [shift = {(0.45, 1)}, rotate=\t*360/11, orange!50!black]
                (0, 0.1) ellipse (0.03 and 0.2);
        }
        \foreach \k in {-13, -11, ..., 13} {
            \pgfmathsetmacro\y{\k * 0.15} 
            \pgfmathsetmacro\ya{\y + 0.2} 
            \pgfmathsetmacro\yb{\y + 0} 
            \pgfmathsetmacro\yc{\y - 0.1} 
            \pgfmathsetmacro\yd{\y - 0.2} 
            \pgfmathsetmacro\x{(mod(999 + \k*13, 21) - 10) * 0.03}
            \pgfmathsetmacro\xa{\x * (1 + (\ya)^4 * 0.2)}
            \pgfmathsetmacro\xb{\x * (1 + (\yb)^4 * 0.2)}
            \pgfmathsetmacro\xc{\x * (1 + (\yc)^4 * 0.2)}
            \pgfmathsetmacro\xd{\x * (1 + (\yd)^4 * 0.2)}
            \draw [preaction={draw=white, line width=2pt}]
                (\xa, \ya) .. controls (\xb, \yb) .. (\xc, \yc);
            \fill [preaction={draw=white, line width=2pt}]
                (\xd, \yd) circle (2pt);
        }
        \draw [shift = {(1, 0)}]
            (0, 1) -- (0, 0.4)
            (0, 0.2) circle (0.2)
            (0, 0.1) -- +(90:0.2)
            (-0.1, 0.25) to [bend left=40] (0.1, 0.25)
            (0, 0) -- (0, -0.3)
            (-0.1, -0.3) -- (0.1, -0.3)
            (-0.2, -0.4) -- (0.2, -0.4)
            (0, -0.4) -- (0, -1)
        ;
    \end{tikzpicture}
    \caption{
        The nanopore sequencer.
        From left to right:
        Nanopores are installed on a membrane.
        Applying external voltage drives the ions to move through the pores,
        which generates currents we can measure.
        The pores have protein engines attached to them.
        These engines like to move DNA strands step-by-step.
        As the nucleotides block the pathway, fewer ions move,
        and less current would be measured.
        And since A, C, G, and T differ every so slightly in sizes,
        we can determine the blocking letter by
        observing the current drop very carefully.
    }                                                         \label{fig:seq}
\end{figure*}

\section{Current Technology and Problem Statement}            \label{sec:bio}

    A DNA coding scheme necessarily contains two parts:
    DNA synthesis (plus data encoding) and
    DNA sequencing (plus data decoding).
    Both parts produce errors and pose unique coding challenges.
    In this section, we explain the state-of-the-art technologies
    that allow us to manipulate DNA and then
    extract a mathematical model.

\subsection{The Synthesis Part}

    For the purpose of storing data in DNA
    rather than merely modifying genes,
    we are required to synthesize arbitrary ACGT sequences.
    We cannot simply cut and splice existing sequences,
    or else the achievable information rate would be severely limited
    by the number of known sequences.
    Hence, even though researchers have cataloged proteins
    that perform such edits,
    de novo synthesis remains difficult because
    strands must be grown nucleotide by nucleotide.

    Currently, the best approach is photolithographic synthesis
    (Figure~\ref{fig:syn}).
    On a base plate are multiple (say one million) \emph{sites}.
    Each site can grow one\footnote{Sometimes,
    each site grows multiple strands simultaneously.
    Those strands will be completely identical unless errors occur,
    so we simplify the story.} strand of arbitrary content.
    But they are protected by caps by default.

    To extend a site, we remove its protective cap
    and let the next nucleotide (say A) flow in the surroundings.
    Attachment is stochastic:
    A will likely bind to an uncapped site, but not always.
    This leads to deletion errors
    at a small empirical rate (typically $1\%$).
    Insertion errors are far rarer because A carries a cap
    preventing chained attachments.

    After strands extend to around $200$--$300$ nucleotides,
    error rates start rising, and so the process usually terminates here.
    The strands are then detached, amplified, and stored.
    In this work, each strand is assumed to be
    256 nucleotides long (chosen so $\log_2 256$ is an integer).
    
    The crux of the photolithographic approach is
    to mask the light source to target only a subset of sites.
    Other technologies include selective charging\footnote{
    Readers are referred to the animation uploaded by the service provider
    \url{https://www.youtube.com/watch?v=tHQedUXfSCo}} or heating.
    This enables large-scale parallel synthesis, and since there is no
    evident upper limit on site count,
    throughput scales readily in that dimension.

\subsection{The Sequencing Part}

    The sequencing part is also done in parallel
    and there are, generally speaking, two approaches.
    The photonic approach prepares fluorescently tagged
    A, C, G, T building blocks and lets DNA duplicate itself.
    As blocks attach, they emit light of different colors.
    A camera captures those emissions and decodes the sequence.
    This approach is highly accurate (error probability $< 0.1\%$)
    but requires more expensive instrumentation.

    The electric approach uses nanopores embedded in a membrane
    and reads sequences as strands translocate.
    As depicted in Figure~\ref{fig:seq},
    each nanopore consists of a narrow pathway
    and a protein engine that moves DNA in discrete steps
    similar to the escapement\footnote{
    Readers are referred to the animation uploaded by
    the device manufacturer:
    \url{https://www.youtube.com/watch?v=RcP85JHLmnI}} in a clock.
    Without the engine, syncing the data becomes nearly impossible.

    A voltage is applied across the membrane,
    driving ions in the surrounding electrolyte
    and generating picoamp-level currents.
    As strands are translocated,
    each nucleotide partially obstructs the pathway.
    This modulates the current in measurable ways.
    A neural network interprets the current trace as letters,
    with error probability ranging from $0.1\%$ to $1\%$.

\subsection{Problem Statement}

    Combining photolithographic synthesis with
    nanopore sequencing generates three coding challenges.
    \begin{itemize}
        \item Because strands float freely in the pool, they are read as if a
            \emph{permutation channel} randomly permutes strand order.
        \item Because DNA is amplified after synthesis and before reading,
            two pores may read strands with identical content.
            This increases reliability on paper (reads can cross-check),
            but complicates capacity analysis and code design.
        \item Per-letter errors arise during synthesis and sequencing,
            including substitutions, deletions, insertions, and repetition.
    \end{itemize}

    Previous work \cite{WaG25}
    addressed the first two bullets and substitution errors
    and showed capacity achievement in certain parameter regimes.
    In this work, we simplify the first two aspects but add deletion
    errors—the error type both synthesis and sequencing generate.
    We demonstrate that Geno-Weaving remains viable.
    Moreover, it is largely agnostic to error type:
    designs for BSC perform fairly well over BDC.

    The following summarizes the scope of this paper.

    \begin{problem}
        The alphabet is either $\{0, 1\}$ or $\{$A, C, G, T$\}$.
        Each strand is $256$ nucleotides long
        and there are $2^8$--$2^{20}$ strands in total.
        Each strand is sequenced exactly once
        with its index known to the decoder.
        The only errors are deletions with probability $0.1\%$--$10\%$.
        What is the best possible code rate in this regime?
    \end{problem}

    We omit the multi-read aspect of nanopore channels in order
    to enable comparison with a representative ``control group'' scheme.
    The control group concatenates deletion codes
    with Reed--Solomon codes in the most straightforward way possible.
    Multiple reads would introduce ambiguity in defining its decoder.

    We also omit the permutation aspect,
    which can be solved by prefixing each strand with its index.
    This index is about $\log n$ symbols long, where $n$ is
    the number of strands, so the overhead is already nearly negligible.
    Also, both our scheme and the concatenation approach need indexing,
    so the overhead will almost\footnote{
    There is a subtle detail regarding
    if the index is protected (a) by a standalone deletion code,
    (b) by the code protecting the entire strand, or (c) not at all.
    Different code designs will lead to different overheads,
    making comparison tedious.
    Luckily, the beginning part of each strand is widely believed to be
    less error prone, so we decide to ignore this aspect.}
    cancel out.

\section{From Deletion Channel to Concatenation}           \label{sec:concat}

    Synchronization error has been a classic subject in coding theory;
    storage media like magnetic tape used to \emph{slip},
    causing synchronization errors.
    Among channel models for this,
    the binary deletion channel is the most notorious:
    it provides the simplest desynchronization mechanism,
    yet it merging two runs of zeros by deleting the ones in between
    is challenging to revert.

    For these reasons and Levenshtein's result \cite{Lev66}
    (that reduces insertion to deletion),
    research has focused primarily on deletion channels.

    In the DNA data storage era,
    deletion channels resurface because
    both synthesis and sequencing introduce deletions at non-negligible rates.
    We next review code designs
    for deletion channels and discuss how they can
    be used on nanopore channels.

\begin{figure}
    \centering
    \begin{tikzpicture} [yscale=0.8]
        \draw [->] (0, 0) -- (0, 5) node [right] {$r / d \log_q n$};
        \foreach \y in {1, 2, 4, 8} {
            \draw (0, {1 + log2(\y)}) node [left] {\y} -- +(0.1, 0);
        }
        \draw [->] (0, 0) -- (7, 0) node [below] {$d$};
        \foreach \x in {1, ..., 6} {
            \draw (\x, 0) node [below] {\x} -- +(0, 0.1); 
        }
        \draw [->] (1, 1) -- (7, 1)
            node [above left] {asymptotic lower bound};
        \draw [->] (1, 1) -- (2, 2) -- (7, 2)
            node [above left] {implicit construction};
        \draw [->] (2, 2) -- (3, 3) -- (7, 3)
            node [above left] {binary explicit construction};
        \draw [->] (1, 1) -- (2, 3.322) -- (3, 4) -- (7, 4)
            node [above left] {quaternary explicit construction};
        \fill [blue] (2, 3.322) circle (2pt)
            node [above left, scale=0.8] {$(2, 5)$};
    \end{tikzpicture}
    \caption{
        The redundancy $r \coloneqq \ell - \log_q |\BB|$,
        normalized by $d \log_q \ell$,
        as a function of the deletion number $d$.
        See Theorems~\ref{thm:d=1}--\ref{thm:r=5} for details.
    }                                                      \label{fig:scalar}    
\end{figure}
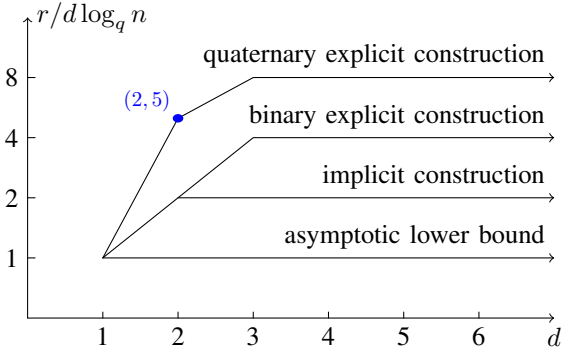

\subsection{Probabilistic Deletion Channels and Codes}

    A probabilistic deletion channel
    can be understood as the following process:
    Given the input $X_1, X_2, \dotsc, X_\ell$,
    it draws a random subset $K \subset [\ell]$,
    the indices of kept symbols.
    For each $s \in [\ell]$, whether $s$ will be in
    $K$ is an independent Bernoulli event with failure rate $\delta$,
    called the \emph{deletion probability}.
    It then outputs $X_{k_1}, X_{k_2}, \dotsc, X_{k_{|K|}}$,
    where $k_j$ is the $j$th smallest element in $K$.

    If $X_1, \dotsc, X_\ell \in \{0, 1\}$, this is called
    a \emph{binary deletion channel}, or BDC$(\delta)$ for short.
    If the alphabet is, say A, C, G, and T, it is
    a \emph{quaternary deletion channel}, or QDC$(\delta)$ for short.
    In the following, we primary talk about BDC
    but all discussion applies to QDC and other alphabets straightforwardly.

    The first question to any channel is existence of capacity.
    More precisely, we ask if there is a limit
    \[ C(\BDC(\delta)) \in [0, 1] \]
    such that one can find codebooks $\BB \subset \{0, 1\}^\ell$
    of sizes $2^{C\ell - o(\ell)}$ with vanishing error probabilities
    as the block length increases $\ell \to \infty$,
    and we are provably not able to for sizes $2^{C\ell + \Omega(\ell)}$.

    The answer is affirmative for a simple super-additivity argument:
    If we have a good codebook $\BB \subset \{0, 1\}^\ell$,
    we can define a larger codebook with codewords of the form
    $u000\dotsm000v$, where $u, v \in \BB$
    and the number of zeros in between is about $\log \ell$.
    This gadget guarantees that the optimal $\log |\BB|$ is roughly\footnote{
    It is not strictly-speaking super-additive because the zero buffer
    increases the block length by $\log \ell$.
    Luckily, this term is sub-linear.}
    super-additive in $\ell$.

    The problem with super-additivity is that
    every time there is a better design than $u000\dotsm000v$,
    the slope will increase.
    But without trying all possible code designs
    we cannot quantify the increment.
    This argument cannot tell us the exact capacity.

    There are, nevertheless,
    provable upper and lower bounds and numerical estimates.
    Readers are referred to \cite{RuC23} for the latest update on this topic.
    For our paper, we only need the following snippet of knowledge.

\begin{figure}
    \centering
    \tikzset{
        A/.style={blue, fill=.!10},
        C/.style={teal, fill=.!10},
        G/.style={blue!30!teal, fill=.!10},
        T/.style={blue!70!teal, fill=.!10},
        base/.style={
            transform shape, above right,
            circle, draw, fill, inner sep=1pt,
        }
    }
    \pgfmathdeclarerandomlist{ACGT}{{A}{C}{G}{T}}
    \begin{tikzpicture}
        \foreach \x in {1, ..., 8} {
            \tikzset{xshift = \x * 1cm}
            \foreach \y in {1, ..., 5} {
                \pgfmathrandomitem\A{ACGT}
                \tikzset{yshift = \y * 0.55cm}
                \draw [\A]
                (0, 0.5) rectangle (0.1, 0) node [base] {\A} (0.5, 0.4);
            }
            \draw [rounded corners=3pt]
                (-0.1, 0.4) rectangle (0.6, 3.4) (0.25, 0.4) node
                [below, align=center, scale=0.7] {inner \\ deletion \\ code}
            ;
        }
        \draw [rounded corners=12pt](0.6, -0.7) rectangle (9, 3.6);
        \draw (4.8, -0.7) node [below] {outer Reed--Solomon code};
    \end{tikzpicture}
    \caption{
        The Concatenation code design:
        Each strand is protected by an inner code
        that is a combinatorial deletion code.
        All strands together are protected by an outer code
        that is a Reed--Solomon code.
    }                                                      \label{fig:concat}
\end{figure}
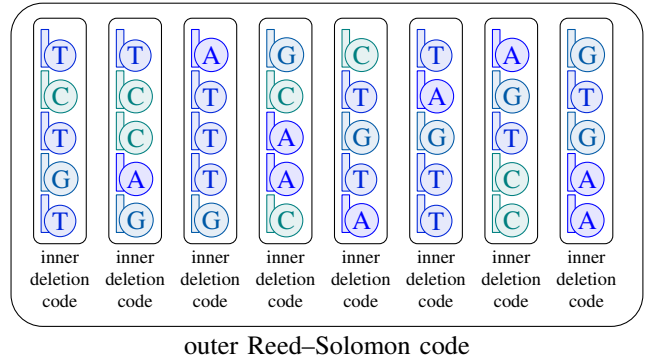

\begin{figure*}
    \includegraphics[width=5.9cm]{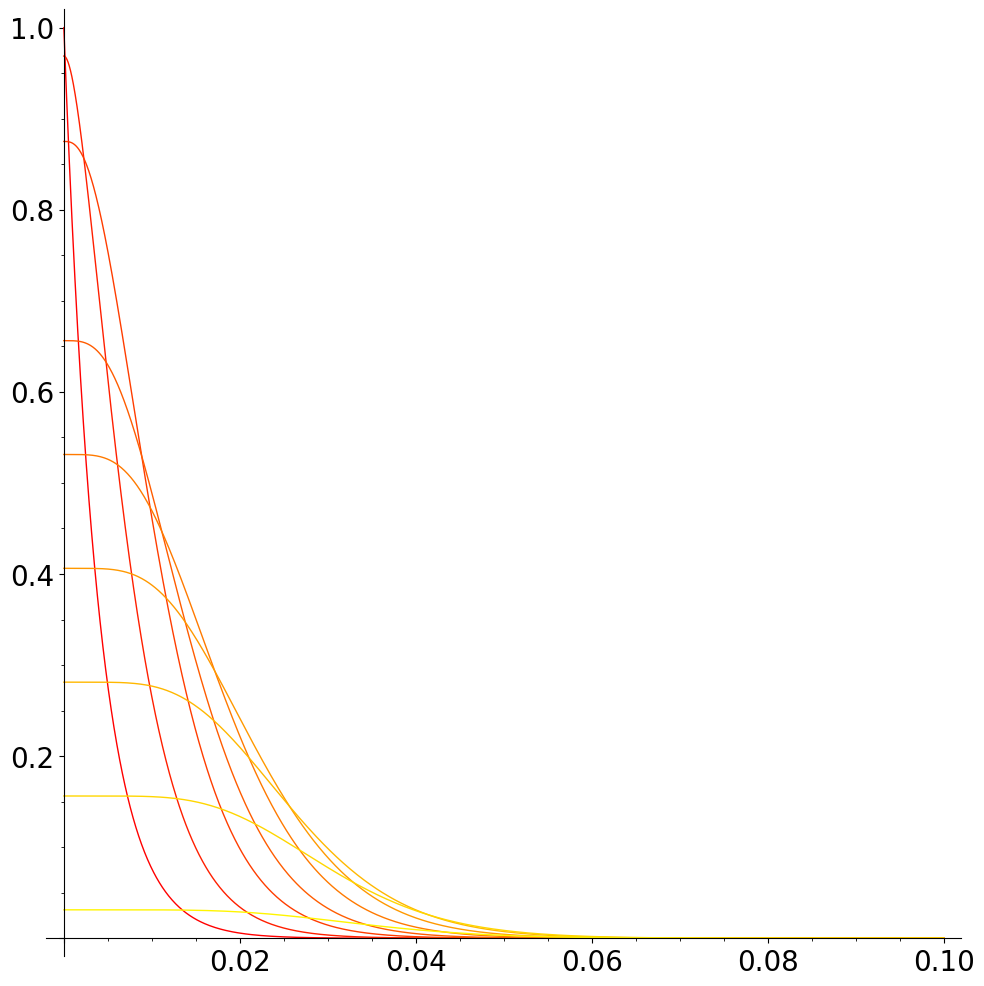}\hfill
    \includegraphics[width=5.9cm]{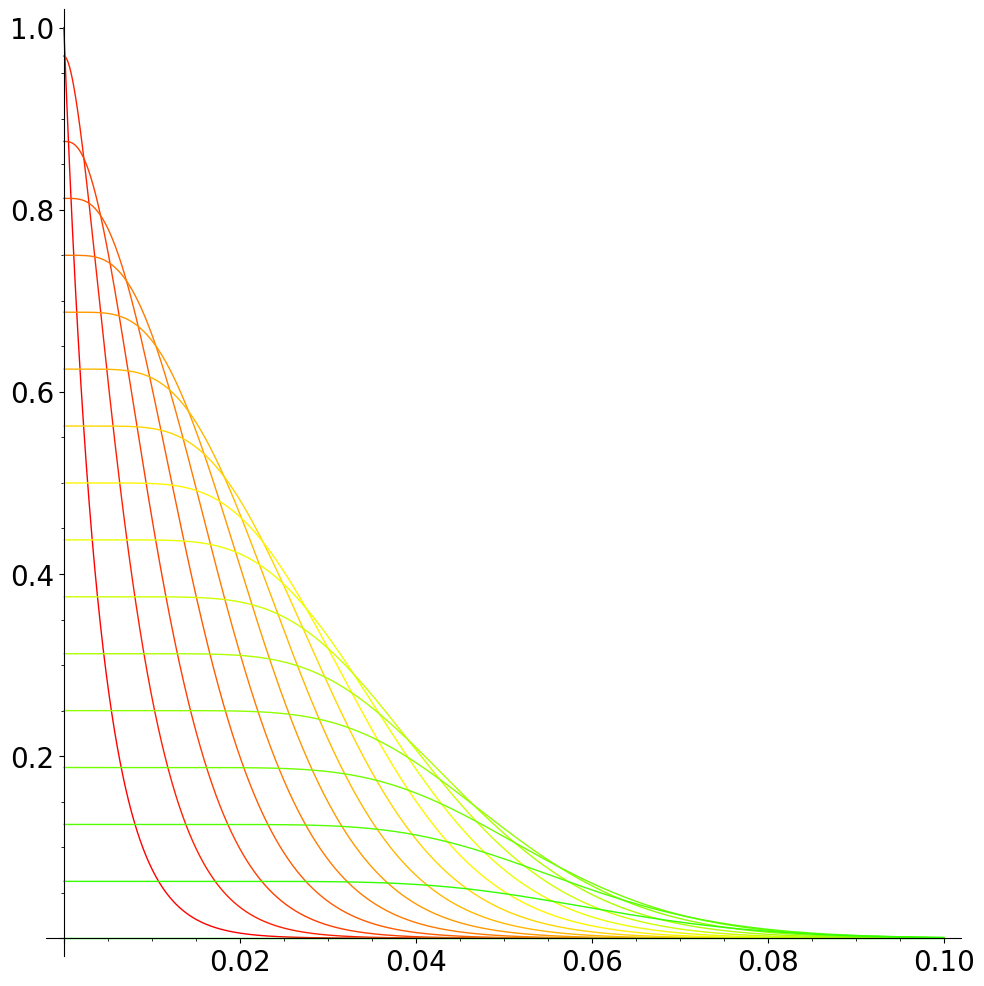}\hfill
    \includegraphics[width=5.9cm]{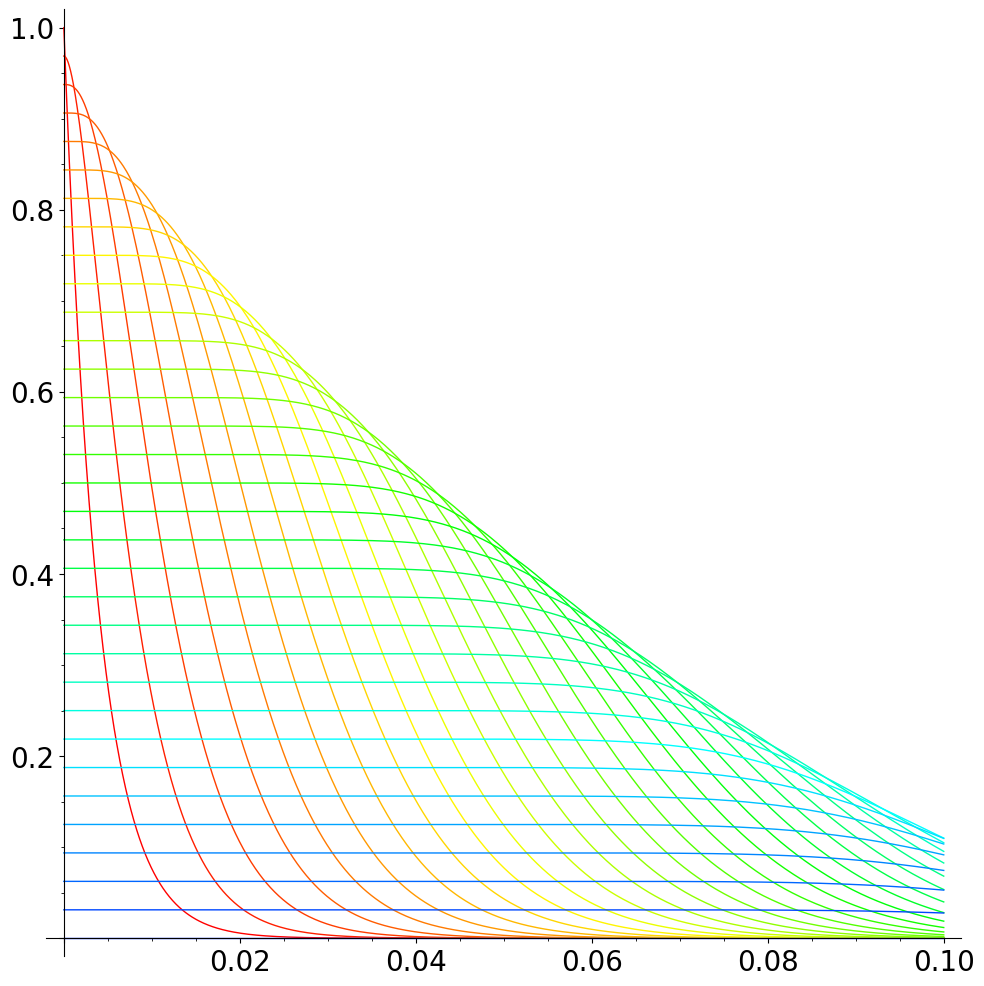}
    \caption{
        The estimated DNA code rate \eqref{concat} if one concatenates
        binary combinatorial deletion codes with Reed--Solomon codes.
        Horizontal axis is deletion probability.
        Vertical axis is code rate.
        each curve represents a different $d$,
        the maximum number of deletions the inner code can handle.
        Lower $d$ has a higher overall code rate at low deletion probability.
        But low-$d$ curves decreases to zero faster than high-$d$ curves
        because the outer codes see more erasures.
        From left to right:
        explicit constructions
        (Theorems~\ref{thm:d=1}, \ref{thm:4d-1}, and \ref{thm:r=4}),
        implicit constructions (Theorem~\ref{thm:dream}),
        and putative constructions that meet the lower bound
        (Theorem~\ref{thm:dream}).
    }                                                     \label{fig:concat2}
\end{figure*}

    \begin{theorem}
        When the capacity is written as a function
        in deletion probability $\delta$
        and $\delta$ is sufficiently small,
        \[ C(\BDC(\delta)) \approx C(\BSC(\delta)) = 1 - h_2(\delta), \]
        where
        \[
            h_2(\delta) \coloneqq
            -\delta \log_2 \delta - (1 - \delta) \log_2 (1 - \delta)
        \]
        is the base-$2$ binary entropy function.
    \end{theorem}

    The reason low-rate BDCs behave like low-rate BSC
    is because decoding a slightly corrupted codeword
    is roughly equivalent to guessing the locations of errors,
    regardless if those are of substitution type or deletion type.
    A supporting evidence of this is \cite{KaD25}, who showed that
    the penalty of low crossover probability and low deletion probability
    are both $h_2(\text{prob})$ and are additive.

    For higher deletion rates, a BDC codeword will contain
    longer runs of zeros and ones.
    The decoder only needs to figure out how many bits are deleted
    in each run, not the exact positions\footnote{
    For example, if $000000$ becomes $00000$,
    the decoder does not have to tell if it is the first zero, the second,
    or the last that is deleted.
    It only has to determine if one bit is deleted from this run.}
    of deletions in their runs.
    Now that the decoder has less uncertainty to resolve,
    the capacity of BDC is strictly higher than that of BSC.
    In particular, BSC$(1/2)$ is already zero capacity
    while $C(\BDC(\delta)) > 0$ for all $\delta < 1$.
    The past two paragraphs suggest that
    using BSC$(\delta)$ to approximate BDC$(\delta)$ is
    (a) accurate for small $\delta$ and (b) pessimistic for all $\delta$.
    
    One final remark regarding probabilistic deletion channels
    is that, although we cannot express its capacity nicely,
    it is still possible to achieve the capacity
    using code constructions with implicit components.
    For instance, in \cite{PLW22},
    the capacity is first approximated by an implicit inner code,
    and then a Reed--Solomon code will boost the error probability
    so that the whole scheme has a positive error exponent.
    In \cite{ArT23}, polar code is used because it can do
    input shaping and noisy channel coding at the same time,
    and the targeted input distribution
    is an implicit Markov chain that achieves the capacity.

    In this work, we do not compare probabilistic deletion codes
    because (a) no explicit parameters are given in the literature
    and (b) their constructions are implicit and not readily implementable
    in practice.
    Instead, we turn into combinatorial deletion codes
    even if our DNA model assumes a probabilistic deletion process.

\subsection{Combinatorial Deletion Channels and Codes}

    A combinatorial deletion channel
    can be understood as the following process:
    Given the input $X_1, X_2, \dotsc, X_\ell$,
    it draws a random subset $K \subset [\ell]$,
    by uniformly choosing $\ell - d$ indices,
    where $d$ is a fixed integer representing the number of deletions.
    It then outputs $X_{k_1}, X_{k_2}, \dotsc, X_{k_{\ell - d}}$,
    where $k_j$ is the $j$th smallest element in $K$.
    Since $d \ge 1 > \delta$, there is no ambiguity;
    we reuse the notation BDC$(d)$ and QDC$(d)$
    for combinatorial deletion channels.

    Note that a combinatorial deletion channel with a fixed $d$
    does not have a capacity in the sense defined previously.
    This is because we can feed a very long codeword
    to amortize the effective deletion rate $\delta \coloneqq d / \ell$.
    Hence the proper measurement of combinatorial deletion channels'
    capability is the co-dimension or the \emph{redundancy}
    \[ r \coloneqq \ell - \log_q |\BB|, \]
    where $q$ is the alphabet size.
    This number is typically a constant multiple of $d \log \ell$.
    For instance, we have the following results.

    \begin{theorem}                                           \label{thm:d=1}
        There are explicit code constructions with redundancy $ \log_q \ell$
        that can correct one deletion.
        See \cite{TeV65} for binary alphabet and
        \cite{Ten84} for general alphabet.
    \end{theorem}

    For more deletions,
    there is a multiplicative gap of $2$.

    \begin{theorem} [\cite{Lev66}]                          \label{thm:dream}
        There are (implicit) codes
        with asymptotic redundancy $2 d \log_q \ell$.
        There are no codes
        with redundancy asymptotically better than $d \log_q n$.
    \end{theorem}

    However, known explicit constructions
    have a higher multiplier at around $4 \log_2 q$.

    \begin{theorem} [\cite{SiB21}]                           \label{thm:4d-1}
        There are explicit binary codes
        with asymptotic redundancy $(4d - 1) \log_2 \ell$.
    \end{theorem}

    \begin{theorem} [\cite{SPC22}]                             \label{thm:4d}
        There are explicit $q$-ary codes
        with asymptotic redundancy $4d \log_2 \ell$.
    \end{theorem}

    Between general $d$ and $d = 1$,
    intermediate results are available for $d = 2$.

    \begin{theorem} [\cite{GuH21}]                            \label{thm:r=4}
        When $d = 2$, there are explicit binary codes
        with asymptotic redundancy $4 \log_2 \ell$.
    \end{theorem}

    \begin{theorem} [\cite{LTX24}]                            \label{thm:r=5}
        When $d = 2$, there are explicit $q$-ary codes
        with asymptotic redundancy $5 \log_2 \ell$.
    \end{theorem}

    See Figure~\ref{fig:scalar} for a visual summary of these theorems.

\begin{figure*}
    \includegraphics[width=5.9cm]{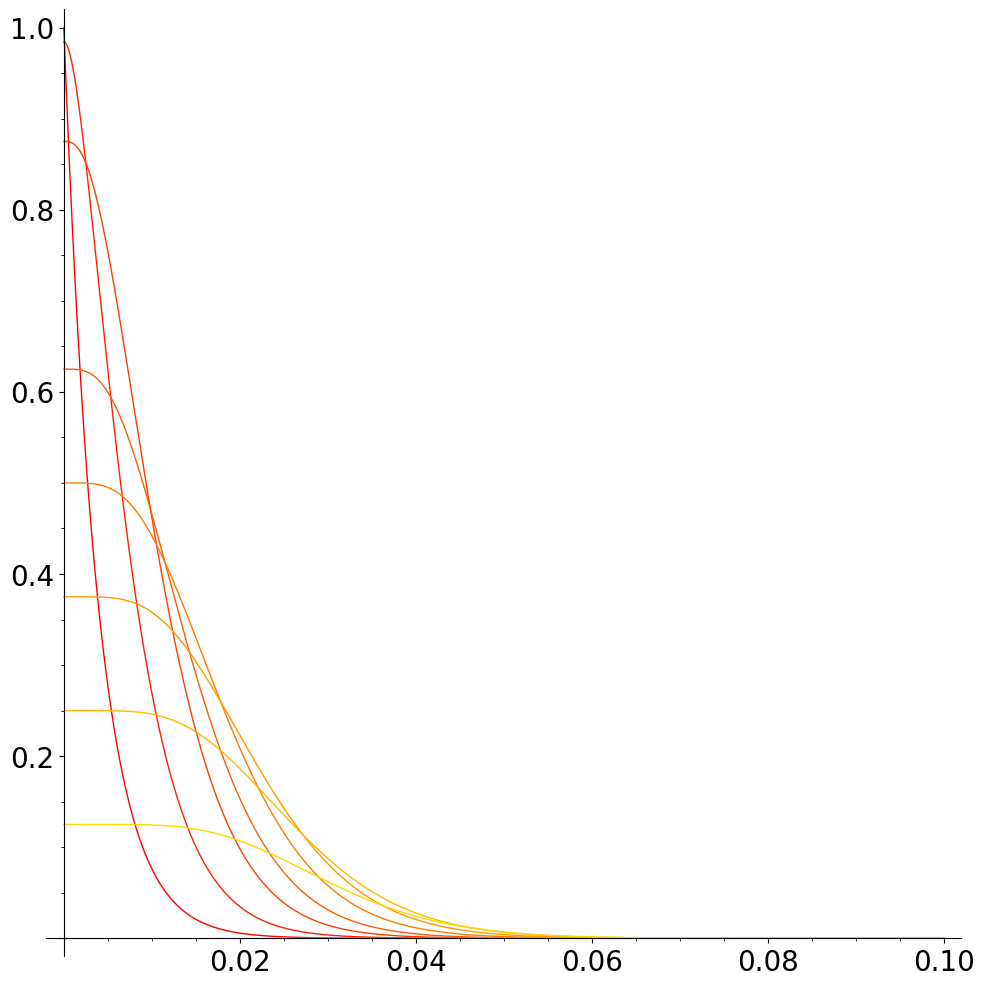}\hfill
    \includegraphics[width=5.9cm]{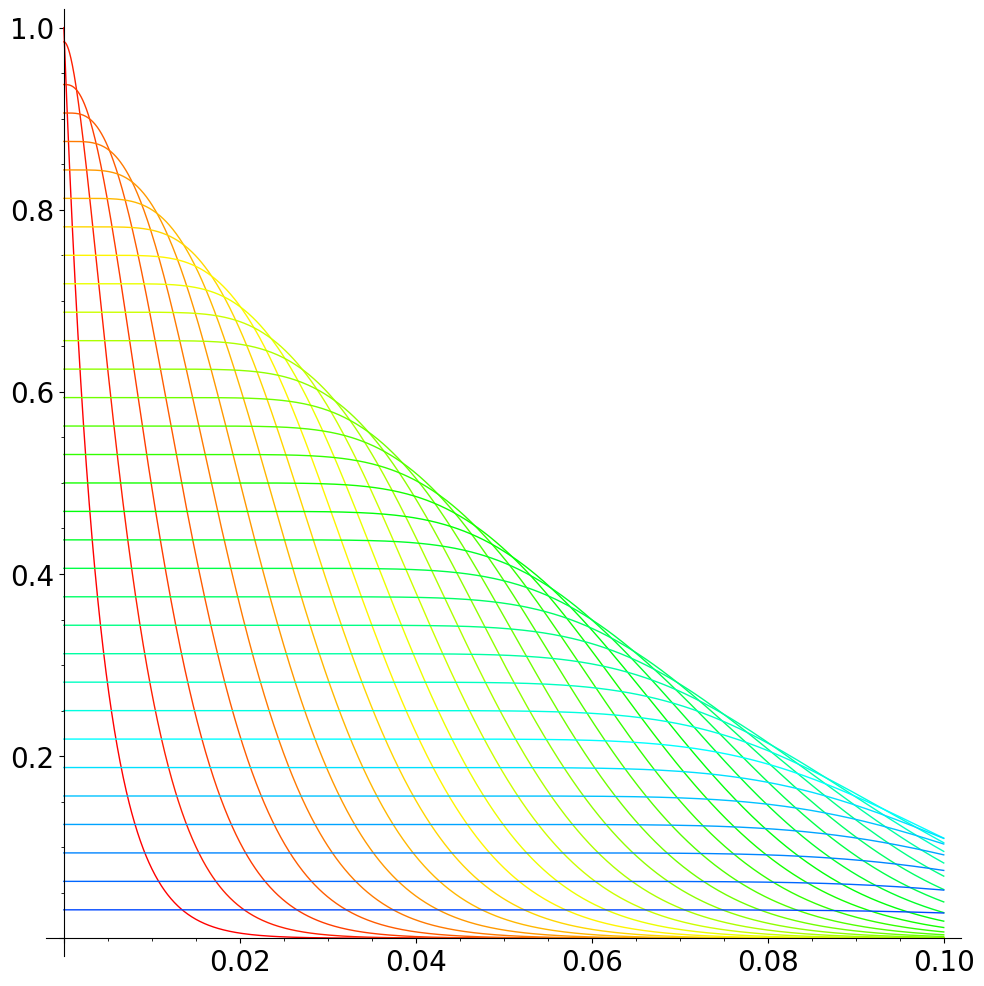}\hfill
    \includegraphics[width=5.9cm]{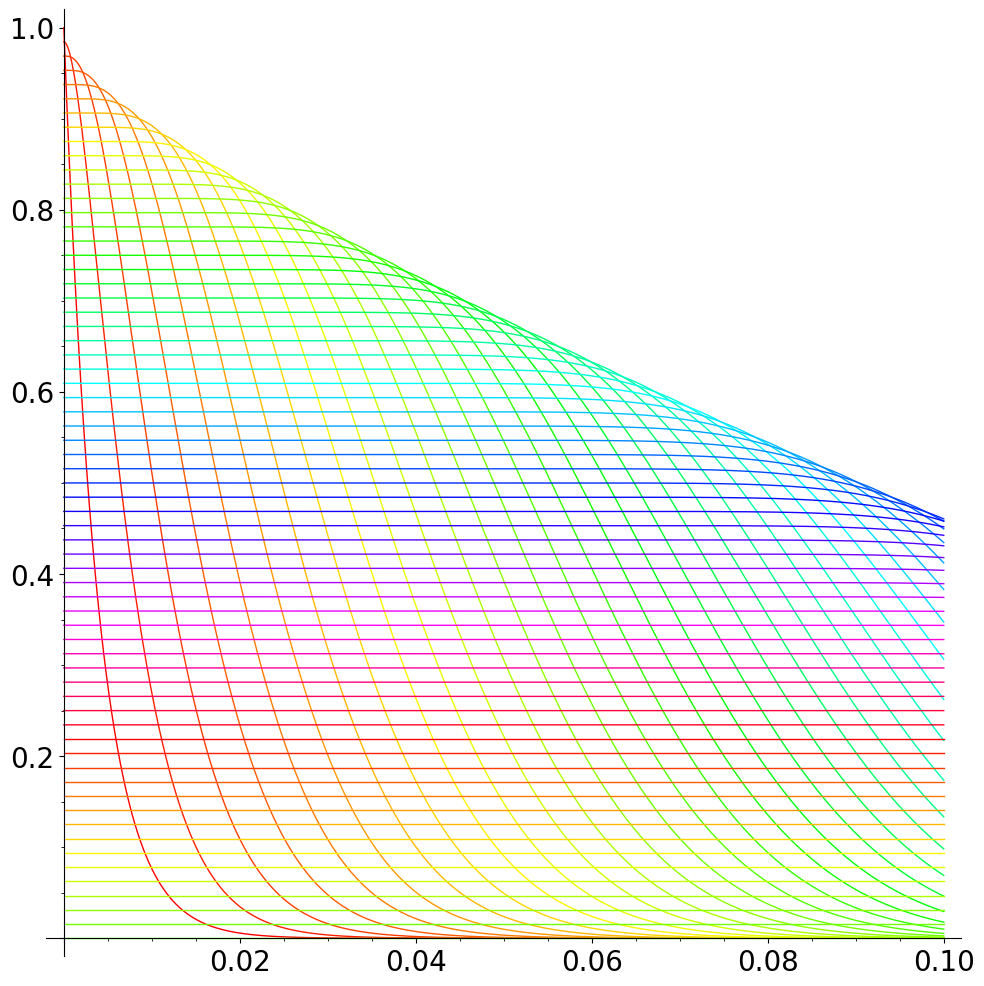}
    \caption{
        The quaternary alphabet version of Figure~\ref{fig:concat2}.
        From left to right:
        explicit constructions
        (Theorems~\ref{thm:d=1}, \ref{thm:4d}, and \ref{thm:r=5}),
        implicit constructions (Theorem~\ref{thm:dream}),
        and putative constructions that meet the lower bound
        (Theorem~\ref{thm:dream}).
    }                                                     \label{fig:concat4}
\end{figure*}

\subsection{Reed--Solomon Codes and Concatenation}

    Recall that probabilistic deletion channels are more natural\footnote{
    For instance, the protein engine on a nanopore slips or
    the letter A refuses to attach to the site in Figure~\ref{fig:syn}.}
    in our setting.
    So the only way to make use of combinatorial deletion codes
    is to choose a $d \approx 256 \delta$ and protect each strand
    as if it will undergo BDC$(d)$ or QDC$(d)$.
    Afterwards, we use an outer Reed--Solomon code to protect the strands
    against erasures, as depicted in Figure~\ref{fig:concat}.

    The key tradeoff lies in the choice of $d$.
    If we choose $d \le 256 \delta$, there will be
    a significant portion of strands that cannot be decoded.
    If we choose $d \gg 256 \delta$,
    the redundancies will penalize the code rate.
    This leads to the following optimization problem.

    Given $\delta$, the number of deletions out of $256$ nucleotides is
    a random variable following the binomial distribution Bin$(256, \delta)$.
    The probability of a strand having $\le d$ deletions is
    $f_\delta(d)$ where $f_\delta$ is the cdf of Bin$(256, \delta)$.
    Suppose we use a Reed--Solomon code with rate $f_\delta(d)$
    to protect the strands, and suppose this outer code always\footnote{
    Note that the RS code might not succeed because the fraction of
    uncorrectable strands fluctuates around its mean $f_\delta(d)$.
    However, the theme of this work is that the number of strands
    is large enough so that the law of large numbers kicks in.
    Also recall that concatenation is used as the control group
    to reflect the advantage of Geno-Weaving.
    This means that we can overestimate the performance
    of RS codes and still prove our point.} succeeds. 
    Then the overall code rate becomes
    the product of inner and outer code rates:
    \begin{equation}
        \Bigl(1 - \frac{r(d)}{256}\Bigr) \cdot f_\delta(d),    \label{concat}
    \end{equation}
    where $r(d)$ is the redundancy as a function in $d$.
    We plot \eqref{concat} in Figure~\ref{fig:concat2} for binary alphabet
    and Figure~\ref{fig:concat4} for quaternary alphabet,
    using Figure~\ref{fig:scalar} to determine $r(d)$.

    For a fixed $d$, it can be seen from Figure~\ref{fig:concat2} that
    the overall code rate is $1 - r(d) / 256$ at $\delta = 0$
    and remains flat for small $\delta$.
    However, beyond a threshold, $f_\delta(d)$ quickly drops to zero
    and so \eqref{concat} also collapses.
    This suggests switching to a larger $d$,
    which starts from a lower initial rate $1 - r(d)/256$
    but remains flat over a wider range of $\delta$.

    Later in Figures \ref{fig:all2} and \ref{fig:all4}, we assume that
    the optimal $d$ (as a function of $\delta$) is always chosen.
    Then, the overall rate of concatenation
    is the maximum of \eqref{concat} over all $d$.

\section{From Polar Codes to Geno-Weaving}                  \label{sec:weave}

    In this section, we discuss how Geno-Weaving provides
    an alternative approach to DNA coding under deletion errors.
    The idea is to protect $X^{(1)}_p, X^{(2)}_p, \dotsc, X^{(n)}_p$
    using one polar code of length $n$,
    repeating this for each position $p \in [256]$.
    Here, $X^{(s)}$ is the $s$th strand
    and $X^{(s)}_p$ is the $p$th letter of the strand.
    We focus on binary alphabet,
    as polar codes over quaternary alphabet \cite{STA09}
    are harder to implement but conceptually similar.

\subsection{Polar Codes}

    Polar codes are a family of capacity-achieving codes \cite{Ari09}.
    One of their features is the flexibility in code rate:
    So long as block length $n$ is a power of $2$,
    explicit constructions exist for every dimension $k \in [n]$.
    Not only that, selecting $k$ can also be done post-simulation,
    i.e., we are essentially simulating for all $k$ at once.
    These two features make polar codes a great tool to study
    the best possible code rate in a scenario like ours.

    To prepare a polar code, first generate
    $n$ \emph{synthetic channels} $W_1, W_2, \dotsc, W_n$
    out of the underlying channel, say BSC$(\delta)$.
    They are also called \emph{bit channels}.
    Fix a bit channel $W_j$, let random variables
    $U_j$ and $V_j$ be its input and output.
    Assume that $U_j \in \{0, 1\}$ is uniform.
    The \emph{equivocation} of $W_j$ is the uncertainty of $U_j$ given $V_j$:
    \[ H(W_j) = H(U_j|V_j) = \EE[h_2(\rho_j(V_j))] \]
    where $\rho_j(v) \coloneqq P(U_j{=}0 \;|\; V_j{=}v)$
    is the posterior probability and $h_2$ is the binary entropy function.
    We can estimate equivocation by sampling $W_j$ $1000$ times
    and take average
    \[
        H(W_j) \approx 1 -
        \frac{1}{1000} \sum_{\sigma=1}^{1000} h_2(\rho_j(V_j^{(\sigma)})),
    \]
    where $V_j^{(\sigma)}$ is the $\sigma$th sample of $V_j$.

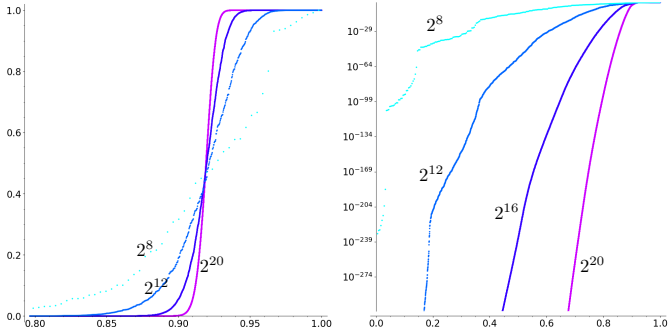
\begin{figure}
    \includegraphics[width=4.4cm]{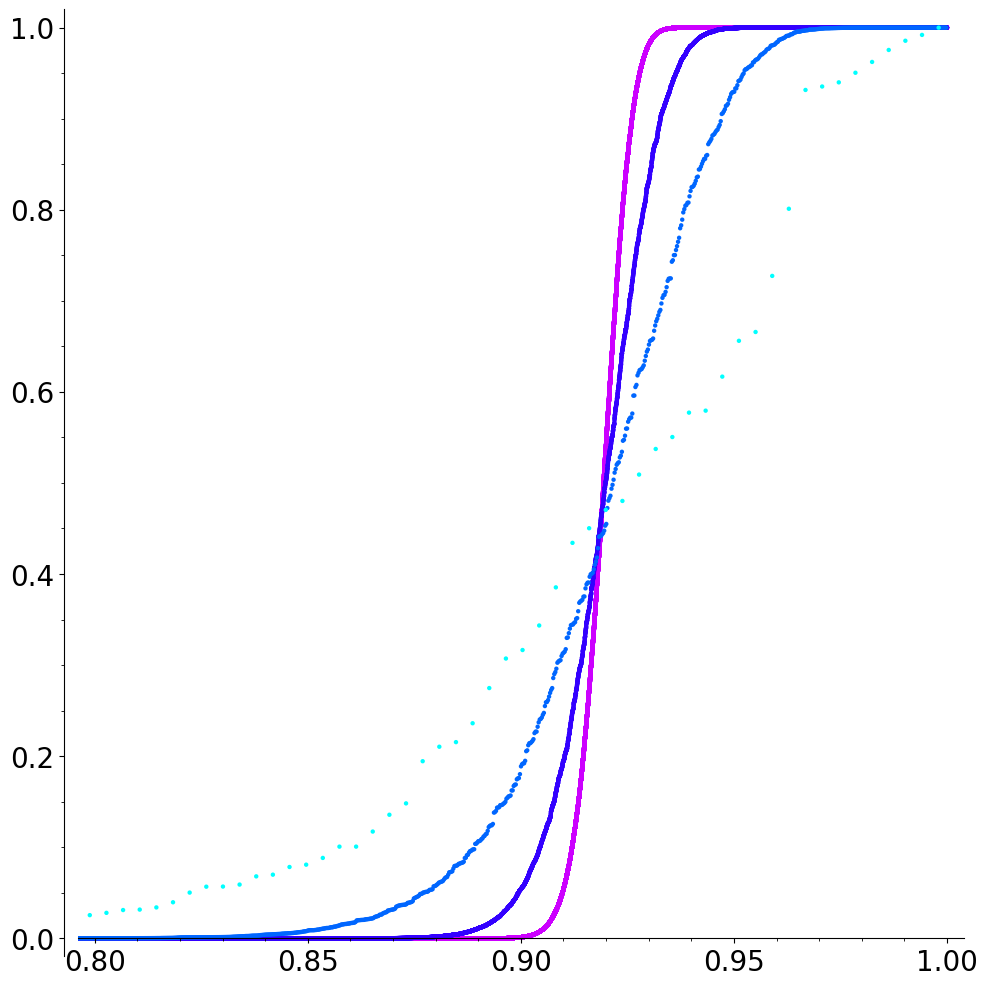}\tikz [overlay] {
        \tikzset{every node/.style = {scale=0.7, inner sep=0}}
        \draw (-1.8, 1) node [below right] {$2^{20}$};
        \draw (-2.2, 0.5) node [above left] {$2^{12}$};
        \draw (-2.4, 1) node [above left] {$2^{8}$};
    }\hfill
    \includegraphics[width=4.4cm]{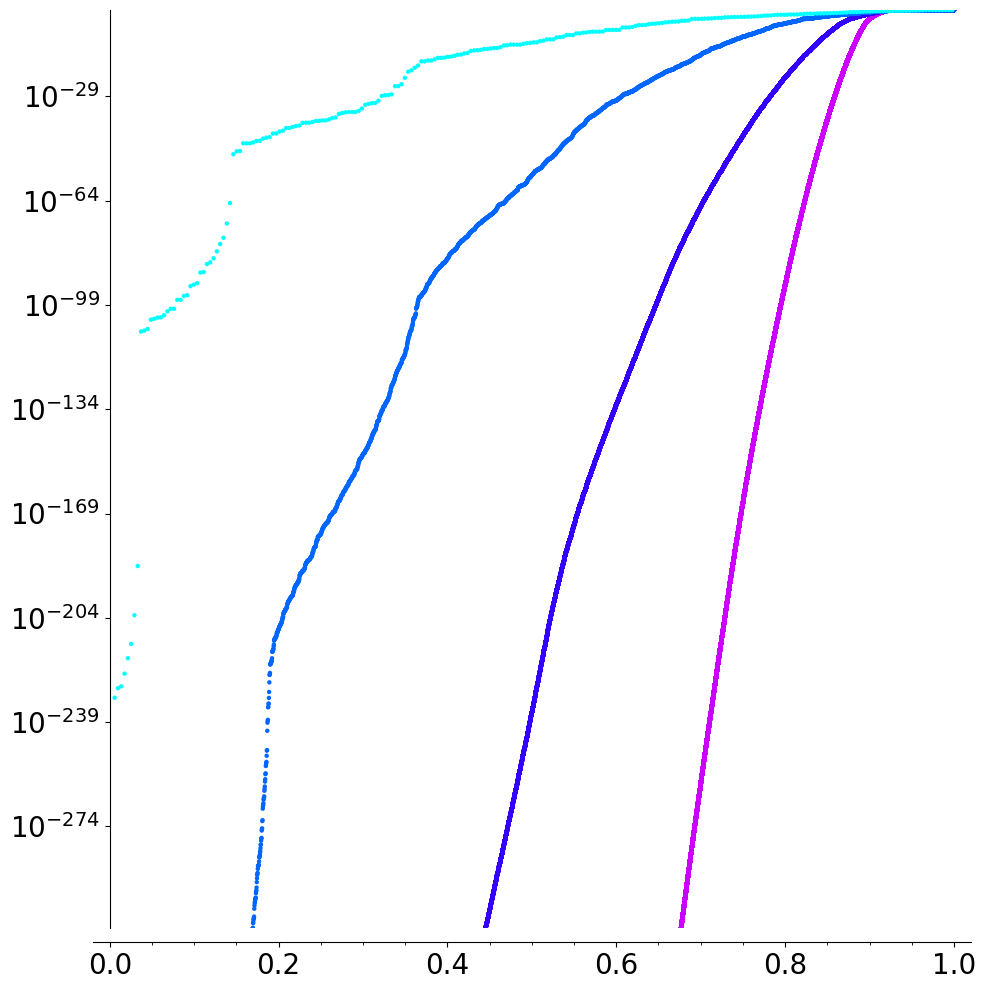}\tikz [overlay] {
        \tikzset{every node/.style = {scale=0.7, inner sep=0}}
        \draw (-1.2, 1) node [below right] {$2^{20}$};
        \draw (-2, 1.5) node [above left] {$2^{16}$};
        \draw (-3, 2) node [above left] {$2^{12}$};
        \draw (-3, 4) node [above left] {$2^{8}$};
    }
    \caption{
        The distribution  of equivocations of bit channels.
        Horizontal axis is bit channel indices
        (after sorting) divided by block length.
        Vertical axis is equivocation estimated by simulating 1000 blocks;
        left is linear and right is semi-log;
        the latter stops at $10^{-300}$ as it is the lower limit
        of double-precision floating-point numbers.
        From cyan (least polarized) to purple (most polarized) are
        block lengths $2^8$, $2^{12}$, $2^{16}$, $2^{20}$.
        The underlying channel is BSC$(1\%)$.
    }                                                       \label{fig:equiv}
\end{figure}

    The name of polar codes comes from \emph{channel polarization}
    \cite{Ari09}, the phenomenon that most equivocations $H(W_j)$
    are either very close to $0$ or very close to $1$.
    See Figure~\ref{fig:equiv} for
    the distribution of empirical equivocations.
    Larger block lengths yield stronger polarization.
    After sorting bit channels by equivocation, the \emph{info set}
    $\II$ is chosen to contain those with the lowest equivocations.
    The bit channels in $\II$ will be used to transmit information.
    The \emph{frozen set} is the complement of the info set,
    and their inputs will be fixed to known values (say all zeros).
    The code rate is $|\II| / n$.

    A rule of thumb is that $\II$ should contain
    bit channels with equivocations less than $1/n$.
    This implies the total uncertainty about all info bits
    is less than $\sum_{j\in\II} H(W_j) < n|\II| \le 1$.
    If one needs better block error probabilities,
    lower the cutoff to, say, $1/10n$.
    In the next subsection,
    we choose $\II$ to be those with equivocations $< 1/256n$,
    so that total uncertainty over all $256$ positions is less than $1$.

\subsection{Geno-Weaving}

    Geno-weaving is the technique developed in \cite{WaG25} to saturate
    the capacity bound developed by \cite{WeM22} and the references therein.
    It consists of two parts.
    \begin{itemize}
        \item $X^{(1)}_p, X^{(2)}_p, \dotsc, X^{(n)}_p$ is protected
            by a block code, for any position index $p \in [256]$.
        \item The strand index is protected by a rateless code
            that spans all positions.
    \end{itemize}
    In this work, we omit the second bullet (strand indexing).
    Therefore, there are $256$ polar code blocks per DNA pool,
    one for each position index $p$.

    The polar codes are designed for BSC$(\delta)$;
    the info set $\II$ contains
    those bit channels with equivocations $< 1/256n$ under BSC$(\delta)$.
    This scheme carries $256|\II|$ information bits
    over $256n$ letters per DNA pool.
    The overall code rate is $|\II| / n$.
    For $n = 2^8, 2^{12}, 2^{16}, 2^{20}, \infty$,
    the code rates as a function in $\delta$
    are plotted in Figure~\ref{fig:all2}.
    The case $n = \infty$ is simply the capacity
    of the underlying channel
    \[ C(\BSC(\delta)) = 1 - h_2(\delta)\]
    because polar codes provably achieve capacity.

    Already we see that at deletion probability $\delta = 1\%$,
    using length-$2^{8}$ polar codes designed for BSC$(1\%)$,
    the overall code rate exceeds that of concatenation.
    For larger block lengths and larger deletion probabilities,
    the advantage grows; even hypothetical optimal deletion codes
    cannot match Geno-Weaving's rate.

    In the next section, we will test this design against deletion errors.

\subsection{Quaternary Alphabet}

    So far, this section focuses on a binary alphabet.
    The story is similar for a quaternary alphabet.
    The only difference is that QDC behaves like a
    quaternary symmetric channel (QSC) rather than a BSC.

    The equivocation of QSC
    with substitution probability $\delta$ is
    \[
        h_4(\delta) \coloneqq
        -\delta \log_4(\delta/3) - (1 - \delta) \log_4 (1 - \delta)
    \]
    and its capacity is
    \[ C(\QSC(\delta)) = 1 - h_4(\delta). \]
    This curve is plotted in Figure~\ref{fig:all4},
    together with the envelops of curves from Figure~\ref{fig:concat4}.

    Comparing Figure~\ref{fig:all4}'s $1 - h_4(\delta)$
    to Figure~\ref{fig:all2}'s $1 - h_2(\delta)$,
    Geno-Weaving now enjoys an even more profound advantage.
    On surface, this is because the capacity of QSC is higher\footnote{
    One needs $\delta = 0.75$ to drive the capacity to $0$.}
    than that of BSC.
    But the bigger issue is with Figure~\ref{fig:scalar}:
    Explicit codes are twice as bad compared to the binary case,
    despite the implicit bound suggesting they should be equally good.
    This again highlights the benefits of
    bypassing detailed deletion-channel study.

    In fact, we can even bypass studying quaternary polar codes.
    We can identify $\{$A, C, G, T$\}$ with $\{0, 1, i, 1+i\}$, which
    decomposes a QDC into the \emph{real} part and the \emph{imaginary} part.
    Each part is now a BDC with the same deletion probability.
    This means that all binary discussion in this paper applies to
    quaternary alphabet straightforwardly.
    In particular, the finite-length curves in Figure~\ref{fig:all2}
    can be reused for Figure~\ref{fig:all4},
    showing the same advantage over concatenation schemes.

\begin{figure}
    \centering
    \tikzset{
        A/.style={blue, fill=.!10},
        C/.style={teal, fill=.!10},
        G/.style={blue!30!teal, fill=.!10},
        T/.style={blue!70!teal, fill=.!10},
        base/.style={
            transform shape, above right,
            circle, draw, fill, inner sep=1pt,
        }
    }
    \pgfmathdeclarerandomlist{ACGT}{{A}{C}{G}{T}}
    \begin{tikzpicture}
        \foreach \y in {1, ..., 5} {
            \tikzset{yshift = \y * 0.6cm}
            \draw [rounded corners=3pt]
                (0.8, -0.05) rectangle (8.7, 0.4)
                node [below right, scale=0.8] {polar};
            ;
        }
        \foreach \x in {1, ..., 8} {
            \tikzset{xshift = \x * 1cm}
            \foreach \y in {1, ..., 5} {
                \pgfmathrandomitem\A{ACGT}
                \tikzset{yshift = \y * 0.6cm}
                \draw [\A]
                (0, 0.5) rectangle (0.1, 0) node [base] {\A} (0.5, 0.4);
            }
        }
    \end{tikzpicture}
    \caption{
        Geno-Weaving's design:
        each position is protected by a polar code.
        There is no outer code;
        the pool fail if any of the position fails.
    }                                                       \label{fig:weave}
\end{figure}
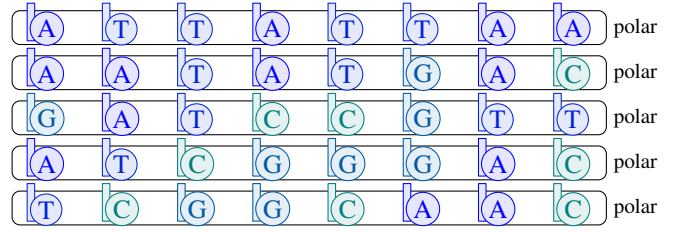

\section{Simulation Results}                                  \label{sec:sim}

    Recall that, in the previous section, we argued that
    Geno-Weaving's code rate exceeds that of concatenation schemes.
    However, a high-rate code without a small pool error probability
    is not useful in practice.
    To prove Geno-Weaving useful, our simulation is set up as follows.

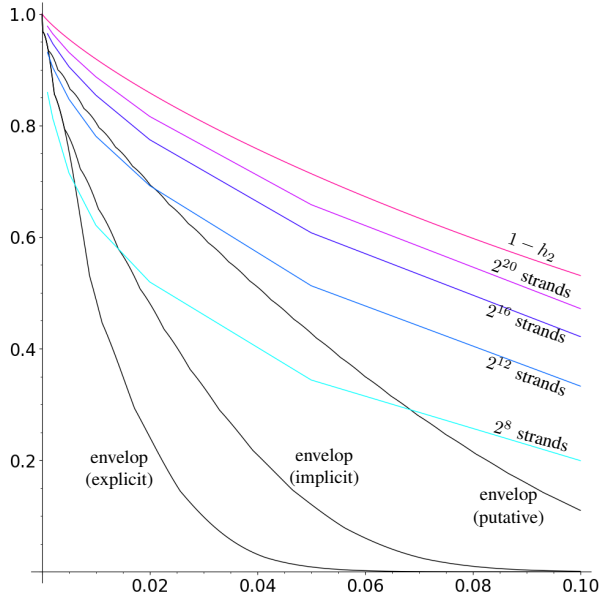
\begin{figure}
    \centering
    \includegraphics[width=8cm]{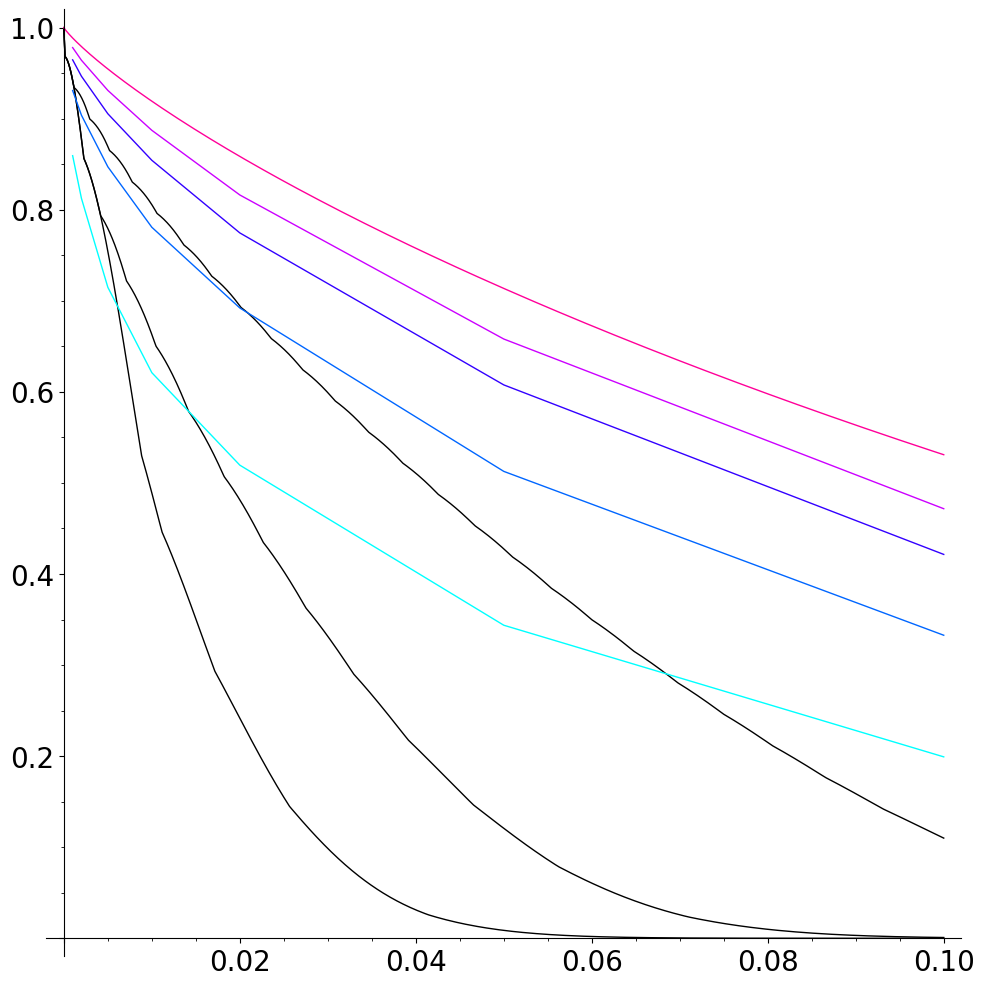}\tikz [overlay] {
        \tikzset{every node/.style = {scale=0.7, inner sep=0}}
        \draw (-6, 2) node [below left, align=center]
            {envelop \\ (explicit)};
        \draw (-4.2, 1.5) node [above right, align=center]
            {envelop \\ (implicit)};
        \draw (-0.8, 1.5) node [below left, align=center]
            {envelop \\ (putative)};
        \draw (-1, 4.6) node [above, rotate=-20] {$1 - h_2$};
        \draw (-1, 4.2) node [above, rotate=-20] {$2^{20}$ strands};
        \draw (-1, 3.8) node [below, rotate=-20] {$2^{16}$ strands};
        \draw (-1, 3.1) node [below, rotate=-20] {$2^{12}$ strands};
        \draw (-1, 2.1) node [above, rotate=-15] {$2^{8}$ strands};
    }
    \caption{
        The estimated code rates for binary alphabet.
        Horizontal axis is deletion/crossover probability.
        Vertical axis is code rate.
        Colorless lines are the \emph{envelopes}
        (the outline of a family of curves)
        of the overall code rates in Figure~\ref{fig:concat2};
        they interpret the horizontal axis as deletion probability.
        Top curve is the capacity of BSC,
        the limit when the number of strands tends to infinity.
        Below it are curves corresponding to block lengths
        $2^{20}$, $2^{16}$, $2^{12}$, and $2^{8}$.
        They interpret the horizontal axis as crossover probability.
    }                                                        \label{fig:all2}
\end{figure}

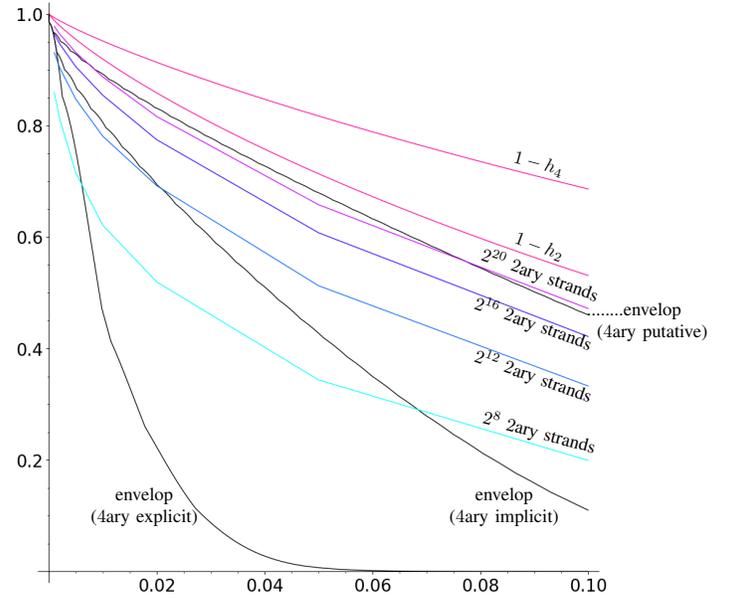
\begin{figure}
    \centering
    \includegraphics[width=8cm]{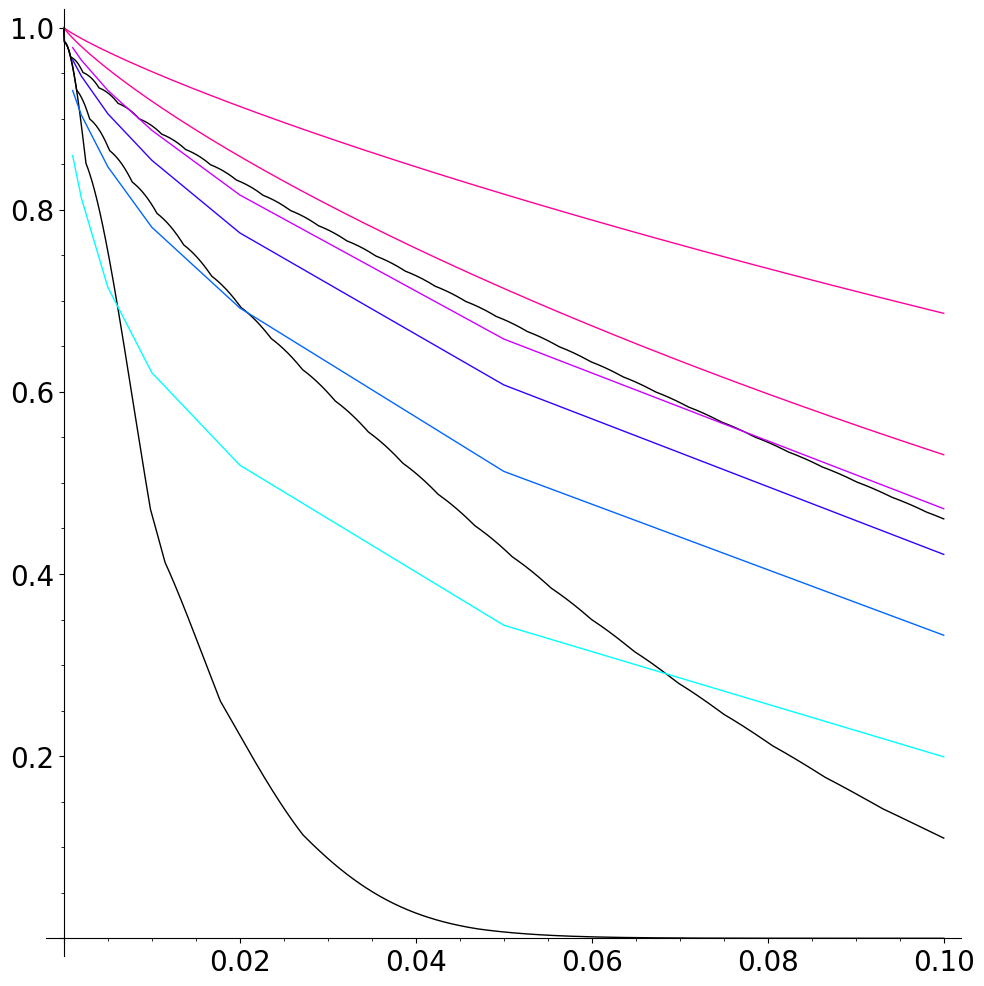}\tikz [overlay] {
        \tikzset{every node/.style = {scale=0.7, inner sep=0}}
        \draw (-5.5, 1.5) node [below left, align=center]
            {envelop \\ ($4$ary  explicit)};
        \draw (-0.7, 1.5) node [below left, align=center]
            {envelop \\ ($4$ary  implicit)};
        \draw 
            (-0.2, 3.7) node [right, align=center]
            {\llap{........}envelop \\ ($4$ary putative)};
        \draw (-1, 5.7) node [above, rotate=-15] {$1 - h_4$};
        \draw (-1, 4.6) node [above, rotate=-20] {$1 - h_2$};
        \draw (-1, 4.2) node [above, rotate=-20] {$2^{20}$ $2$ary strands};
        \draw (-1, 3.8) node [below, rotate=-20] {$2^{16}$ $2$ary strands};
        \draw (-1, 3.1) node [below, rotate=-20] {$2^{12}$ $2$ary strands};
        \draw (-1, 2.1) node [above, rotate=-15] {$2^{8}$ $2$ary strands};
    }
    \caption{
        The comparison between QDC codes
        and binary polar code with the real--imaginary decomposition.
        Colorless lines are the envelops of the overall code rates
        in Figure~\ref{fig:concat4}.
        Top curve is the capacity of QSC.
        Other colored curves are the copied from Figure~\ref{fig:all2}.
    }                                                        \label{fig:all4}
\end{figure}

    To construct polar codes, $256000$ blocks are sampled.
    Block length is $n \in \{2^8, 2^{12}\}$.
    The underlying channel is BSC$(\delta)$;
    $\delta \in \{0.1\%$, $0.2\%$, $0.5\%$, $1\%$, $2\%$, $5\%$, $10\%\}$.
    The bit channels with empirical equivocations $< 1/256n$
    are used as info bits.
    The code rates thus obtained
    coincide with those in Figure~\ref{fig:all2}.

    To test performance over a realistic nanopore channel,
    each letter is deleted with probability $\delta$.
    Each strand has exactly $256$ letters before deletion;
    strands are padded with erasure symbols after deletions.
    A pool contains $n$ strands.
    Then $1000$ pools are sampled.
    At every position $p \in [256]$, the polar decoder is applied.
    If the decoder makes a mistake, the whole pool is declared a failure.
    The numbers of failures are in Table~\ref{tab:del}.

    We now explain how the polar decoder discovers deletion:
    Suppose that the first position $p = 1$ is the all-zero codeword.
    Suppose that all but one observation,
    say $Y^{(99)}_1$ at the 99th strand, is $1$.
    Seeing an all-zero but one-one pattern,
    the decoder immediately concludes that
    that $X^{(99)}_1$ should be $0$.
    This contradicts the observation $Y^{(99)}_1 = 1$.
    So it infers that $X^{(99)}_1 = 0$ was deleted,
    and $Y^{(99)}_1$ is actually\footnote{
    Or it could be the case that
    $Y^{(99)}_1$ is taken from $X^{(99)}_3$
    because both $X^{(99)}_1$ and $X^{(99)}_2$ are deleted.
    However, we only consider small deletion probability,
    and so it is less unlikely that two deletions occur.}
    taken from $X^{(99)}_2 = 1$.

    We then \emph{push} the observed strand $Y^{(99)}$ as follows.
    \[
        \def\se{\kern-1.3em\searrow\kern1.3em}
        \begin{array} {llllllllll}
            Y^{(99)}_1 & Y^{(99)}_2 & Y^{(99)}_3
            & \cdots & \cdots & Y^{(99)}_{255} & Y^{(99)}_{256}\\
            & \se & \se & \se
            &  & \se & \se \\
            0 & Y^{(99)}_1 & Y^{(99)}_2
            & \cdots & \cdots & Y^{(99)}_{254} & Y^{(99)}_{255} \\
        \end{array}
    \]
    As a result, the second block the polar decoder sees is
    \[
        Y^{(1)}_2, \dotsc, Y^{(98)}_2, Y^{(99)}_1,
        Y^{(100)}_2, \dotsc, Y^{(n)}_2.
    \]
    In general, pushing is performed every time
    the observation does not match its decoded value.
    Therefore, what the $p$th block the decoder sees
    consists of $Y^{(s)}_{p-d(s,p)}$ for $s \in [n]$, where $d(s,p)$
    is the number of deletions the $s$th strand found before position $p$.

\subsection{Pull for Insertions}

    One can see that the pushing mechanism generalizes to
    a \emph{pulling} mechanism that handles insertions.
    More precisely, if the decoded value contradicts
    the observation $Y^{(99)}_1$,
    it must be the case that $Y^{(99)}_1$ is an inserted value.
    We infer that $X^{(99)}_1$ appears at $Y^{(99)}_2$,
    so we pull the observations as below.
    \[
        \def\se{\kern-1.3em\swarrow\kern1.3em}
        \begin{array} {llllllllll}
            Y^{(99)}_1 & Y^{(99)}_2 & Y^{(99)}_3
            & \cdots & \cdots & Y^{(99)}_{255} & Y^{(99)}_{256}\\
            & \se & \se & \se
            &  & \se & \se \\
            Y^{(99)}_2 & Y^{(99)}_3 & Y^{(99)}_4
            & \cdots & \cdots & Y^{(99)}_{256} & Y^{(99)}_{257} \\
        \end{array}
    \]
    As a result, the second block the polar decoder sees is
    \[
        Y^{(1)}_2, \dotsc, Y^{(98)}_2, Y^{(99)}_3,
        Y^{(100)}_2, \dotsc, Y^{(n)}_2.
    \]
    In general, pulling is performed every time
    the observation does not match its decoded value.
    Therefore, what the $p$th block the decoder sees
    consists of $Y^{(s)}_{p+i(s,p)}$ for $s \in [n]$, where $i(s,p)$
    is the number of insertions the $s$th strand found before position $p$.

    We run the same simulation as before,
    but now with insertions instead of deletions.
    The numbers of erroneous pools are in Table~\ref{tab:ins}.

    Because codes resisting $d$-deletions resist $d$-insertions as well
    \cite{Lev66}, Figure~\ref{fig:all2} remains a valid comparison
    between the concatenation approach and Geno-Weaving
    when insertions are considered instead of deletions.
    Moreover, repetition is just insertion
    that always coincides with the previous letter,
    so Figure~\ref{fig:all2} remains a valid comparison
    when the underlying channel produces repetition errors.

\subsection{Mixed Error Types}

    Readers might have realized that the pushing and pulling mechanisms
    exploit the assumption of single error type, which implies that
    the observations can be resynchronized by simply shifting
    in a fixed direction until they match the decoded values.

    This, however, breaks down if any two of
    substitution, insertion, and deletion occur simultaneously.
    In such cases, we must fall back to estimating the posterior probability
    \[
        P(X^{(99)}_p{=}0 \;|\; X^{(99)}_1 \dotsm X^{(99)}_{p-1}
        Y^{(99)}_1 \dotsm Y^{(99)}_{256+i-d}),
    \]
    where $256+i-d$ is the final length of the observations.
    We believe this is possible using a trellis decoder
    with $O(p^2)$ complexity.
    We do not like this and are still working on better solutions
    that strike a balance between performance and complexity.

\section{Conclusion}

    We take Geno-Weaving designed for substitution errors
    and apply it to the deletion case.
    Simulations show that its pool error probability
    is low enough for practical use.
    We also discuss concatenation schemes
    and upper-bound their code rates.
    Geno-Weaving operates at rates
    exceeding what concatenation schemes can possibly achieve.

    The apparent source of advantage is that
    deletion channels are difficult to work with.
    The performance of a concatenation scheme is
    limited by what explicit codes are available today.
    Geno-Weaving bypasses this by designing for substitution channels.
    
    The more fundamental source of advantage is that,
    even with hypothetical bound-matching deletion codes,
    the short strand length ($200$--$300$ nucleotides)
    imposes a strong finite block length penalty.
    By coding in the orthogonal direction that scales up more easily,
    Geno-Weaving enjoys a \emph{block length gain}.

\bibliographystyle{IEEEtran}
\bibliography{NanoGain-32}

\begin{table}
    \centering
    \caption{
        The number of erroneous DNA pools out of 1000 pools.
        Columns are deletion probabilities.
        Rows are the number of strands per pool.
    }                                                         \label{tab:del}
    \begin{tabular} {cccccccc}
        \toprule
        ~        & 0.1\% & 0.2\% & 0.5\% & 1\% & 2\% & 5\% & 10\% \\
        \midrule
        $2^{12}$ & 3     & 9     & 14    & 14   & 6  & 3   & 5    \\
        $2^8$    & 5     & 9     & 8     & 8    & 11 & 5   & 5    \\
        \bottomrule
    \end{tabular}
\end{table}

\begin{table}
    \centering
    \caption{
        The number of erroneous DNA pools out of 1000 pools.
        Columns are insertion probabilities.
        Rows are the number of strands per pool.
    }                                                         \label{tab:ins}
    \begin{tabular} {cccccccc}
        \toprule
        ~        & 0.1\% & 0.2\% & 0.5\% & 1\% & 2\% & 5\% & 10\% \\
        \midrule
        $2^{12}$ & 8     & 6     & 5     & 9    & 7  & 5   & 1    \\
        $2^8   $ & 12    & 13    & 6     & 7    & 9  & 8   & 0    \\
        \bottomrule
    \end{tabular}
\end{table}

\end{document}